%% file: Final_version.tex
\documentclass[journal,english]{IEEEtran}

\usepackage{epsfig}
\usepackage{times}
\usepackage{float}
\usepackage{afterpage}
\usepackage{amsmath}
\usepackage{amstext,cite}
\usepackage{amssymb,bm}
\usepackage{latexsym}
\usepackage{color}
\usepackage{graphicx}
\usepackage{amsmath}
\usepackage{amsthm}
\usepackage{graphicx}
\usepackage[center]{caption}
\usepackage{pstricks}
\usepackage{caption}
\usepackage{subcaption}
\usepackage{booktabs}
\usepackage{multicol}
\usepackage{lipsum}% dummy text
\usepackage[T1]{fontenc}
\usepackage{hyperref}
 \usepackage{aecompl}
 \usepackage{mathrsfs}

 \usepackage{arydshln}

\usepackage{soul}
\usepackage{cancel}

\usepackage{changes}

\allowdisplaybreaks

\input{macros}

\usepackage{amsmath}
\usepackage{tikz}

\begin{document}

\title{On the Optimal Memory-Load Tradeoff of Coded Caching for Location-Based Content}
\iffalse
\author{
%
Kai~Wan,~\IEEEmembership{Member,~IEEE,}
Minquan~Cheng,
Mari~Kobayashi,~\IEEEmembership{Senior~Member,~IEEE,}
and~Giuseppe Caire,~\IEEEmembership{Fellow,~IEEE}
%

\thanks{
K.~Wan and G.~Caire are with the Electrical Engineering and Computer Science Department, Technische Universit\"at Berlin, 10587 Berlin, Germany (e-mail:  kai.wan@tu-berlin.de; caire@tu-berlin.de). The work of K.~Wan and G.~Caire was partially funded by the European Research Council under the ERC Advanced Grant N. 789190, CARENET.}
\thanks{
M. Cheng is with Guangxi Key Lab of Multi-source Information Mining $\&$ Security, Guangxi Normal University,
Guilin 541004, China  (e-mail: chengqinshi@hotmail.com). The work of Cheng was in part supported by NSFC (No. 62061004), Guangxi Collaborative Innovation Center of Multi-source Information Integration and Intelligent Processing, and the Guangxi Bagui Scholar Teams for Innovation and Research Project.}

\thanks{
M.~Kobayashi  is with the Technical University of Munich, Germany (email: mari.kobayashi@tum.de).}
}
\fi
\author{
Kai~Wan,~\IEEEmembership{Member,~IEEE,}
Minquan~Cheng,
Mari~Kobayashi,~\IEEEmembership{Senior~Member,~IEEE,}
and~Giuseppe Caire,~\IEEEmembership{Fellow,~IEEE}

\thanks{
K.~Wan and G.~Caire are with the Electrical Engineering and Computer Science Department, Technische Universit\"at Berlin, 10587 Berlin, Germany (e-mail:  kai.wan@tu-berlin.de; caire@tu-berlin.de). The work of K.~Wan and G.~Caire was partially funded by the European Research Council under the ERC Advanced Grant N. 789190, CARENET.}
\thanks{
M.~Cheng is with Guangxi Key Lab of Multi-source Information Mining $\&$ Security, Guangxi Normal University,
 China  (e-mail: chengqinshi@hotmail.com).  The work of M.~Cheng was    in part supported by NSFC (No.62061004, U21A20474), Guangxi Collaborative Innovation Center of Multi-source Information Integration and Intelligent Processing, the Guangxi Bagui Scholar Teams for Innovation and Research Project, and the Guangxi Talent Highland Project of Big Data Intelligence and Application.
}
\thanks{
M.~Kobayashi is with the Department of Electrical and Computer Engineering, Technical University of Munich, 80333 Munich, Germany (e-mail: mari.kobayashi@tum.de).
}
}
\maketitle
%\IEEEpeerreviewmaketitle{}

\begin{abstract}
Caching at the wireless edge nodes is a promising way to boost the spatial and spectral  efficiency, for the sake of     alleviating networks from content-related traffic.
 Coded caching originally introduced by Maddah-Ali and Niesen  significantly speeds up communication efficiency  by transmitting multicast messages simultaneously useful to multiple users.
Most prior works on coded caching are based on the assumption that each user may request all content  in the library.   
%Due to the requirement on the development of the Internet of Vehicles (IoV) facing to the ever-increasing vehicular applications (e.g., autonomous vehicles, intelligent transportation, in-car entertainment, etc.), this paper formulates the coded caching problem for location-based content downloads in the vehicular networks with edge cache nodes.
However, in many applications the users are interested only in a limited set of content   that depends on their location. For example, %visitors in a museum may stream audio and video related to the artworks in the room they are visiting,  or 
  assisted self-driving vehicles may access super High-Definition maps of the area through which they are travelling.  Motivated by these considerations, this paper formulates the coded caching problem for location-based content   with edge cache nodes.  
The considered problem includes a content server  with access to $\Nsf$ location-based files (e.g., High-Definition maps),
$\Ksf$ edge cache nodes located at different regions, and $\Ksf$ users (i.e., vehicles) each of which is in the serving region of  one cache node and can retrieve the cached  content of this cache node with negligible cost. Depending on the    location, each user   only requests a file from a location-dependent subset of the library.
The objective is to minimize  the worst-case    load (i.e., the worst-case number of broadcasted bits from the content server   among all possible demands). 
For this novel coded caching problem, we propose a highly non-trivial converse bound under uncoded cache placement (i.e., each cache node directly copies some library bits in its cache), which shows that a simple achievable scheme is optimal under uncoded cache placement. In addition, this achievable scheme is also proved to be generally order optimal within a factor of $3$. Finally, we extend the coded caching problem for location-based content   to the multiaccess coded caching topology originally proposed by Hachem et al., where each user is connected to $\Lsf$ nearest cache nodes. When $\Lsf \geq 2$, we characterize the exact optimality on  the worst-case load.
\end{abstract}


\section{Introduction}
\label{sec:intro}
Caching reduces peak traffic by taking advantage of devices' memories distributed across the network to duplicate content during off-peak hours, such that
the network traffic is shifted from peak to  off-peak hours.  A caching system is operated in two phases: i) {\it placement phase:} each user stores some bits in
its cache without knowledge of later demands; ii) {\it delivery phase:} after each user has made its request and
according to cached content, the server transmits packets in order to satisfy the user demands. The goal is to
minimize the transmission load such that user demands can be satisfied.

Information theoretic coded caching was originally proposed by Maddah-Ali and Niesen (MAN) in~\cite{dvbt2fundamental} for a shared-link caching 
system where a server with a library of $\Nsf$ equal-length files is connected to $\Ksf$ users through a noiseless shared link and each user can store $\Msf$ files in its local cache.
%system  containing a server with a library of $\Nsf $ equal-length files, which is connected to $\Ksf$ users through a noiseless shared link, each of which can store $\Msf$ files in their local cache. 
   Each user demands an arbitrary file in the library during the delivery phase. The MAN scheme uses a combinatorial design in the placement phase such that each multicast message transmitted during the delivery phase simultaneously satisfies the demands of multiple users.
Under the constraint of uncoded cache placement (i.e., each user directly caches a subset of the library bits) and for the worst-case load among all possible demands, the MAN scheme was proved to be optimal when $\Nsf\geq \Ksf$~\cite{indexcodingcaching2020,exactrateuncoded}. 
Provided  the observation that some MAN linear combinations are redundant if there exist files demanded by several users, the authors in~\cite{exactrateuncoded} improved the MAN delivery scheme and achieved the optimal worst-case load under the constraint of uncoded cache placement for any $\Ksf$. It was also proved in~\cite{yufactor2TIT2018} that the multiplicative gap between the optimal caching scheme with uncoded cache placement and any caching scheme with coded cache placement is at most $2$.

Caching at the wireless edge nodes   reduces both the backhaul traffic
and the transmission time for high-volume data delivery~\cite{FemtoCaching,LiuWirelesscaching}.
% Compared to the end-user-caches, which are heavily limited by the storage size  of each user's device (e.g., a library size is usually much larger than the storage size of a mobile device) and only useful to  one user, the caches at edge devices  can be made much larger and can be accessed by multiple users.
  Although extensively considered in the literature, the end-user-caches, e.g. mobile devices,  have some limitations as they have typically small storage size (compared to the library size) and are useful only for one user device. By contrast, the caches at edge devices, e.g. small base stations, have larger storage capability and can be accessed by multiple users. 
In addition, location-based content could be placed into the edge caching devices at different locations, such that when the mobile clients enter one area, they can retrieve the location-based content  for this area from the corresponding edge caching devices.

%Coded caching strategies have also been extended to various edge caching systems.

\paragraph*{Coded caching    for location-based content}
 \begin{figure}
    \centering
 \includegraphics[scale=0.3]{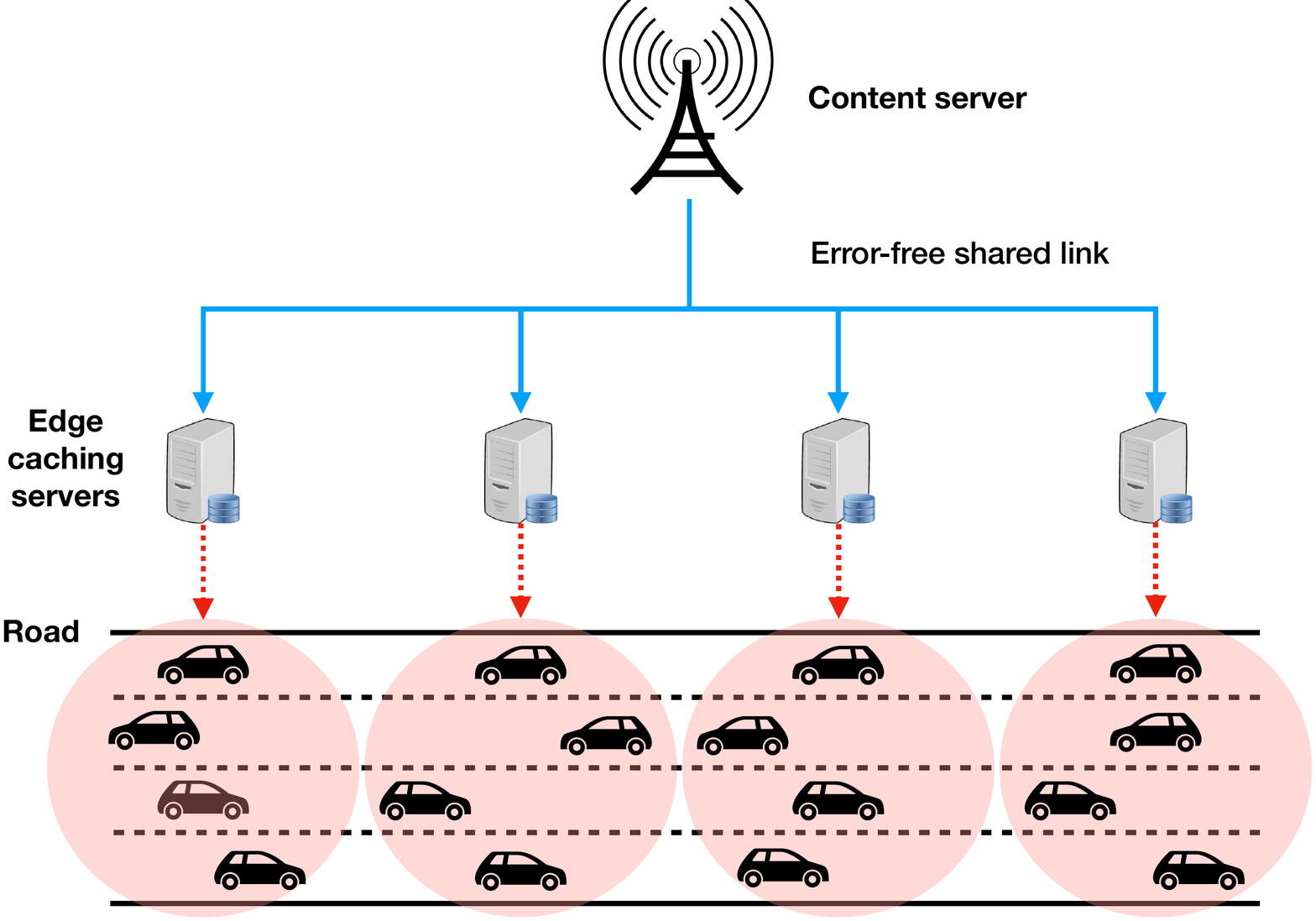}
\caption{\small  Vehicular network  coded caching  problem  for location-based content.}
\label{fig:VNET}
%\vspace{-5mm}
\end{figure}

 Recently the emerging vehicular applications (such as autonomous vehicles, intelligent transportation, high-quality Internet navigation and entertainment, etc.) bring a revolutionary change to the traditional vehicular transportation, and meanwhile lead  to a  dramatically increasing number of demands on data services.
To scope with the large volumes of data, edge caching was widely used in  the vehicular networks, to list a subset of literature~\cite{roadsize2017,NDNvehicle,delayminized2017,mobilityawre2019,mobilityprediction2018,heterogeneousinfocontent2019,locationbased2020,codedcachinginIOV2021}.
%In this paper, we focus on the coded caching problem for location-based content   in the vehicular networks with edge cache nodes  illustrated in  Fig.~\ref{fig:VNET},
  Motivated by this type of applications, in this paper we focus on the coded caching problem for location-based content. 
In order to keep the problem tractable and nevertheless provide some fundamental insight, we consider a very simplistic model of vehicular network with edge cache nodes as illustrated in Fig.~\ref{fig:VNET}, including one content server with access to $\Nsf$  location-based files such as High-Definition (HD) maps  and/or location-based  advertising, entertainment, and services. 
Following the original MAN coded caching model, we assume that each content file has equal size of $\Bsf$ bits.  
The content server is connected to $\Ksf$ edge cache nodes through an error-free shared link. We assume that each cache node  is in a fixed assigned location and has a local cache with size
$\Msf\Bsf$ bits; for example, the cache nodes could be   roadside units (RSUs),  mobile edge caching (MEC)  servers, unmanned aerial
vehicles (UAVs) hovering on assigned geographic areas.
Each cache node is accessible to the vehicles in one area without load cost;\footnote{\label{foot:no cost} %The fact that the edge cache nodes can be accessed locally, can be motivated by the fact that for example they use a local link (e.g., very wideband mmWave proximity link) of very high capacity.  
 The cost-free access between edge caches and users can be justified by dedicated local links  of very high capacity (e.g. wideband mmWave proximity links).  
Models of this kind are widely assumed in the literature of coded edge caching systems such as 
FemtoCaching~\cite{FemtoCaching2013TIT}, multiaccess coded caching~\cite{Hachem2017multiaccess}, and coded caching with shared-caches~\cite{parrinello2018sharedcache}.}
  that is, we  count only  the transmission load from the content server.
The whole road modelled as  a ring is divided into $\Ksf$ non-overlapping regions of equal size, each of which is
 connected to one cache node.\footnote{\label{foot:ring} The ring networks are   very   popular and widely studied model in the literature, such as circular Wyner model for interference networks with  limited interference from neighbours~\cite{uplink2009sandero,complete2016wigger}, 
 multiaccess coded caching~\cite{Hachem2017multiaccess}, etc. A ring network is  interesting in the theoretic sense,  because   the boundary effect at both ends is ignored.}
  Each vehicle in one region demands one location-based  file corresponding to its   region.
The set of possible demanded files in the $k^{\text{th}}$ region is denoted by $\Dc_k$, for each $k\in [\Ksf]$, where 
the set $\Dc_k \subseteq [\Nsf]$ is formed by three subsets: 
%two subsets of the same size denoted by $\asf$  of files also present in the neighbouring regions to the left  and to the right ($\Dc_{k-1}$, and $\Dc_{k+1}$, respectively, where the region index is intended  modulo $\Ksf$ because of the ring topology); 
two subsets  of equal size $\asf$ that represent files also present in the neighbouring regions to the left $\Dc_{k-1}$ and to the right $\Dc_{k+1}$, respectively;  a subset of size $\bsf$ that represent files uniquely present in $\Dc_k$.
%a subset of size $\bsf$ of files uniquely present in $\Dc_k$. 
%the files in $\Dc_k$ are composed of three parts: (i) $\asf$ common files which are also in the left-hand side neighbouring region; (ii) $\bsf$ files which are only in $k^{\text{th}}$ region; (iii) $\asf$   common files which are also in the right-hand side neighbouring region.

\begin{figure}
    \centering
 \includegraphics[scale=0.19]{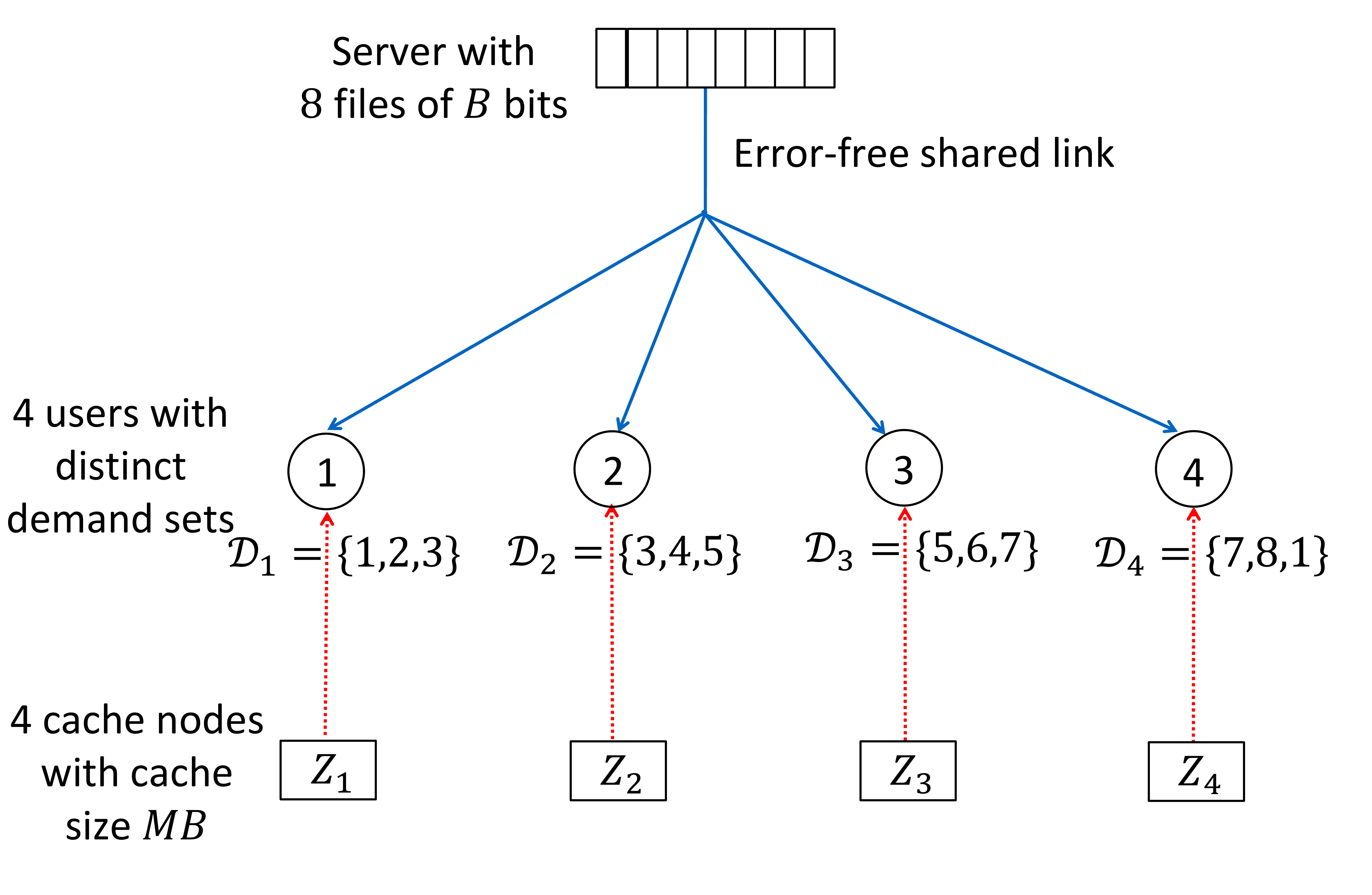}
\caption{\small  The information theoretic model of the considered   coded caching  problem  for location-based content  with $\Ksf=4$, $\Nsf=8$, and $\asf=\bsf=1$.}
\label{fig:VNET math}
%\vspace{-5mm}
\end{figure}

  We treat
the considered problem  as an information theoretic   coded caching problem  with $\Ksf$  users (i.e. vehicles) such that each user is located
  at one distinct region,\footnote{\label{foot:one user one region} In a more practical scenario with multiple vehicles per region,  one direct solution is to divide the whole transmission into multiple rounds, where in each round we serve one user in each region. However, the converse bounds derived in this paper would not directly apply. 
  It is one of our on-going works to characterize the optimal memory-load tradeoff for this case. 
  }  as illustrated in   Fig.~\ref{fig:VNET math}.
In the cache placement phase, each cache node stores some content of the $\Nsf$ files without knowledge of the users' later demands. In the delivery phase, the user at the $k^{\text{th}}$ region  requests a file in $\Dc_k$. According to the users' demands, the central server   broadcasts $\Rsf \Bsf$ bits to all users, such that each user can recover its demanded file from the broadcasted packets and the stored content of its connected cache node. The objective is to  minimize $\Rsf$ for the worst-case demand(s) over the restricted set of possible (location-based) demands.

\paragraph*{Relation to the existing works}
Our considered problem differs from the existing works~\cite{roadsize2017,NDNvehicle,delayminized2017,mobilityawre2019,mobilityprediction2018,heterogeneousinfocontent2019,locationbased2020,codedcachinginIOV2021} for the vehicular networks with edge cache nodes on two aspects: (i) we consider a coded caching problem     for which we need to design the cache placement of the cache nodes and the multicast messages transmitted by the content server to minimize the broadcasted load, while the techniques in~\cite{roadsize2017,NDNvehicle,delayminized2017,mobilityawre2019,mobilityprediction2018,heterogeneousinfocontent2019,locationbased2020}  are based on uncoded caching;  (ii) there exists some overlap on the demand sets of each two neighbouring regions in our problem, which is not considered in~\cite{roadsize2017,NDNvehicle,delayminized2017,mobilityawre2019,mobilityprediction2018,heterogeneousinfocontent2019,locationbased2020,codedcachinginIOV2021}.
Coded caching with location-based content delivery was also considered in~\cite{nonsymmloccaching}. Different from our  edge caching  model, the users in the setting considered in~\cite{nonsymmloccaching} have their own cache memories   and are  mobile    with  uniform access
probability to each location, where
 each location corresponds to one specific file and the placement phase is done without knowing the locations of the users.

Our considered problem can be seen as a special case of the coded caching problem with different demand sets, where the set of possible demanded files by each cache-aided user is different. The   exact optimality results on the memory-load tradeoff were  fully characterized in~\cite{fullheterogeneity2019} for the two-user two-file case, and were partially
 characterized in~\cite{insufficiency2019,overlappingdemand2020} for the two-user $\Nsf$-file case.
 For the general case where the users are divided into classes and the users in the same class have the same demand set, some achievable schemes were proposed in~\cite{heterogeneousprofile2019,averageheteo2020}. Namely, these schemes let  each user use some fraction of its cache to store the common files among different groups and remaining fraction to store the unique files in its group, such that    the MAN   caching scheme  could be used to transmit the common files among different groups and the unique files in each group, respectively. 
Very recently, by focusing on a special case  where there exist the same number of common files in the demand  sets of each $\alpha$-subset of users (e.g. $\alpha=\Ksf$ reduces to the setup of~\cite{dvbt2fundamental}),
 the authors in~\cite{brunero2021unselfish} showed that  the direct use of the  MAN caching scheme regardless the users' demand sets
  can yield unbounded gains over the best selfish caching scheme which lets each user only cache some bits from the files in its demanded files.

 However, except some asymptotic optimality guarantees for some specific cases, the optimality for the coded caching problem with different demand sets is generally open.
In our paper, we focus on a   special structure of different demand sets, where the number of common demanded files by each two neighbouring users
is denoted by $\asf$ and the number of unique demanded files by each user is denoted by  $\bsf$.

\paragraph*{Our Contributions}
In addition to the   formulation of  on the novel coded caching  problem  for location-based content, our main contributions are as follows:
\begin{itemize}
\item By proposing a highly non-trivial converse bound under uncoded cache placement, we prove that the memory sharing among two simple memory-load tradeoff points $(0,\Ksf)$ and $(2\asf+\bsf,0)$, and one achieved tradeoff point by the MAN scheme $\left(\frac{\Nsf}{\Ksf},\frac{\Ksf-1}{2} \right)$, is optimal under uncoded cache placement.
The converse strategy in~\cite{indexcodingcaching2020,exactrateuncoded} for the MAN caching problem  under uncoded cache placement leads to a loose converse in our   problem, because it sums many redundant inequalities in order to derive the final lower bound.
\item  Compared  to a novel  cut-set converse bound, this achievable scheme is proved to be order optimal within a factor of $3$.
\item We   also formulate our location-based content   problem  with   the multiaccess coded caching topology, as illustrated in Fig.~\ref{fig:VNET multiaccess math}.   In this multiaccess coded caching topology originally considered in~\cite{Hachem2017multiaccess}, each user is connected to $\Lsf \geq 2$ neighbouring cache nodes.
 By extending the proposed achievable scheme and the cut-set converse bound to this novel model, we characterize the exact optimality.  It is interesting to see that when $\Lsf \geq 2$,  it does not reduce the load if each user is allowed to access more than $2$ caches. This result provides a very important insight on the design of edge caching schemes for location-based content, showing that 
essentially localized access through high-capacity proximity links 
to the neighbouring caches is indeed sufficient to achieve the optimal load of the (costly) cellular   broadcast channel. 
\end{itemize}

\paragraph*{Paper Organization}
The rest of this paper is organized as follows.
Section~\ref{sec:model} formulates the coded caching  problem  for location-based content   and reviews some related results.
Section~\ref{sec:results} introduces the main results in this paper.
Section~\ref{sec:extension}  extends the proposed bounds to the multiaccess  coded caching  problem  for location-based content.
Section~\ref{sec:conclusion} concludes the paper, while some proofs can be found in the Appendix.

\paragraph*{Notation Convention}
%We use the following notation convention.
Calligraphic symbols denote sets,
bold symbols denote vectors,
%{\red MATRICES?}
and sans-serif symbols denote system parameters.
We use $|\cdot|$ to represent the cardinality of a set or the length of a vector.
Sets of consecutive integers are denoted as
$[a:b]:=\left\{ a,a+1,\ldots,b\right\}$ and $[n] := [1:n]$.
The symbol $\oplus$ represents bit-wise XOR.
%Finally, $\mathbb{E}[\cdot]$ represents the expected value of a random variable;
%$[a]^+:=\max\{a,0\}$; and
$a!=a\times (a-1) \times \cdots \times 1$ represents the factorial of $a$.
$<b>_a$ represents the modulo operation on $b$ with  integer divisor $a$ and in this paper we let $<b>_a\in \{1,\ldots,a \}$ (i.e., we let $ <b>_a=a$ if $a$ divides $b$).
%the number of $k$-permutations of $n, n\geq k,$ is indicated as $P(n,k):=n\cdot(n-1)\cdots(n-k+1)$.
We use the  convention that $\binom{x}{y}=0$ if $x<0$ or $y<0$ or $x<y$.

\section{System Model and Related  Results}
\label{sec:model}

\subsection{System Model}
\label{sub:system}
The information theoretic formulation of the $(\Ksf,\asf,\bsf)$   coded caching  problem  for location-based content   is given as follows, illustrated in Fig.~\ref{fig:VNET math}.
A central server has access to a library of $\Nsf$ location-based files, denoted by $W_1,\ldots,W_{\Nsf}$, each of which contains $\Bsf$ i.i.d. bits.
$\Bsf$ is assumed to be large enough such that any subpacketization on the files is possible.
We consider a one-dimensional cyclic route.  $\Ksf$ cache nodes are distributed on the route where the distance between two neighbouring cache nodes is identical.
Each cache node can cache up to $\Msf \Bsf$ bits.
Each user on the route can retrieve the cached  content from its nearest cache node. Thus the whole route is divided into $\Ksf$ regions, each of which corresponds to one cache node.
Based on the location of the $k^{\text{th}}$ region,
  each user connected to the cache node $k$ is only interested in the files whose indices are in the set  $\Dc_k$.
 Intuitively, $\Dc_k$ is the union of  three disjoint  parts:
 \begin{itemize}
 \item  $\Dc_{k,1}:= \Dc_k \cap \Dc_{<k-1>_{\Ksf}}$, representing the  $\asf$ common files which can also be demanded by the users  in  the $k^{\text{th}}$ region and  in the left-hand side neighbouring region, i.e., the $(<k-1>_{\Ksf})^{\text{th}}$ region.
 \item $ \Dc_{k,2}:=\Dc_k \setminus  \left(\cup_{j\in [\Ksf]\setminus \{k\} } \Dc_j \right)$, representing the  $\bsf$   files which can only be  demanded by the users  in  the $k^{\text{th}}$ region.
  \item  $\Dc_{k,3}:=  \Dc_k \cap \Dc_{<k+1>_{\Ksf}}$, representing the  $\asf$ common files which can also be demanded by the users  in  the $k^{\text{th}}$ region and   in the right-hand side neighbouring region, i.e., the $(<k+1>_{\Ksf})^{\text{th}}$ region.
\end{itemize}
Due to the topology of the one-dimensional cyclic route, there does not exist any  file which can be demanded by two users in  two non-neighbouring regions, i.e., $\Dc_{k_1} \cap \Dc_{k_2} =\emptyset $ where $  <k_1-k_2>_{ \Ksf} \in [2:\Ksf-2]$.  Hence, we have $\Nsf:= \Ksf(\asf+\bsf)$ and
\begin{align}
 \Dc_k&:=  \underbrace{ \left[(k-1)(\asf+\bsf)+1 :  k\asf+(k-1)\bsf\right] }_{:=\Dc_{k,1} }   \nonumber\\& \cup  \underbrace{ \left[k\asf+(k-1)\bsf+1 :  k(\asf+\bsf)\right]  }_{:=\Dc_{k,2} }  \nonumber\\& \cup   \underbrace{  \left[< k(\asf+\bsf)+1 >_{  \Ksf(\asf+\bsf) }:<(k+1)\asf+ k \bsf >_{  \Ksf(\asf+\bsf) } \right]   }_{:=\Dc_{k,3} }. \label{eq:Dk}
\end{align}

 \iffalse
\begin{align}
 \Dc_k&:=  \left[(k-1)(\asf+\bsf)+1 :  k\asf+(k-1)\bsf\right] \cup \left[k\asf+(k-1)\bsf+1 :  k(\asf+\bsf)\right]  \nonumber\\& \cup \left[< k(\asf+\bsf)+1>_{\Ksf(\asf+\bsf)}:< (k+1)\asf+ k \bsf >_{  \Ksf(\asf+\bsf) } \right] , \forall k\in [\Ksf]. \label{eq:Dk1}
\end{align}
 \fi

The server communicates with $\Ksf$ users through an error-free shared link. Each user connected to cache node $k$ can retrieve the content stored in cache node $k$.
 In this paper, we focus on  the communication bottleneck   on the shared link from the server to the users; thus we assume
that each user can retrieve the cached content  from its connected cache node  without any cost.

The system operates in two phases.
\paragraph*{Cache Placement Phase}
 During the cache placement phase, each cache node stores information about the  $\Nsf$  files in its local cache without knowledge of the users' demands. We denote the cached content of cache node $k\in [\Ksf]$ by $Z_k=\phi_k(W_1,\ldots,W_{\Nsf})$, where
 \begin{align}
 \phi_k :  [0:1]^{\Nsf \Bsf} \rightarrow [0:1]^{\Msf \Bsf}, \ k\in [\Ksf].
 \end{align}
Let ${\bf Z}=(Z_1,\ldots,Z_{\Ksf})$ be the cached content of all cache nodes.
%The placement phase is uncoded if the bits of the   files are directly copied into the caches.

\paragraph*{Delivery Phase}
  As explained in Footnote~\ref{foot:one user one region}, we assume that there is exactly one user in each region who makes the request in the delivery phase,
where the user in the $k^{\text{th}}$ region is called user $k$, for each $k\in [\Ksf]$.
The demand vector is defined as ${\bf d}:= (d_1 ,\ldots ,d_{\Ksf} )$, where $d_k \in \Dc_k $ represents to the index of the file demanded by user $k \in [\Ksf]$. The demand vector ${\bf d}$ is known to the server and all users.
Given $({\bf Z}, {\bf d})$, the server broadcasts the message $X =\psi({\bf d},W_{1},\ldots,W_{\Ksf} )$, where
 \begin{align}
\psi: \Dc_1 \times \cdots \times \Dc_{\Ksf} \times    [0:1]^{\Nsf \Bsf} \rightarrow  [0:1]^{\Rsf \Bsf},
 \end{align}
for some non-negative number $\Rsf $ referred to as load.

\paragraph*{Decoding}
 Each user $k  \in [\Ksf]$ decodes its desired file $F_{d_k}=\xi_k({\bf d}, Z_k ,X)$, where
 \begin{align}
\xi_k: \Dc_1 \times \cdots \times \Dc_{\Ksf} \times    [0:1]^{\Msf \Bsf}  \times  [0:1]^{\Rsf \Bsf}   \rightarrow  [0:1]^{\Bsf}, k\in [\Ksf].
 \end{align}
%$\forall k\in [\Ksf].$
\paragraph*{Objective}
 For any cache size $\Msf \in [0,\Nsf]$, we aim to determine the {\it minimum worst-case load} among all possible demands,
defined as the smallest $\Rsf$ such that there exists an ensemble of
placement functions $\phi_k, k\in [\Ksf],$
encoding function $\psi$, and
decoding functions $\xi_k, k\in [\Ksf],$
satisfying all the above constraints.
%~\eqref{eq: encoding function def} and~\eqref{eq: decoding functions def}.
The optimal load is denoted by $\Rsf^{\star}$.

Note that if $\Ksf=1$, we have $\asf=0$ and $\Nsf=\bsf$. The considered problem becomes the $1$-user MAN coded caching problem, where the uncoded caching scheme is optimal. In the rest of this paper, we consider $\Ksf\geq 2$.

 %The optimal load under the constraint of uncoded cache placement is denoted by  $\Rsf_{\rm u}^{\star}$.

\paragraph*{Uncoded Cache Placement}
The cache placement policy is  {\it uncoded} if the bits of the   files are directly copied into the cache nodes.
 Under the constraint of uncoded cache placement, we can partition each file $W_{i}$ where $i\in [\Nsf]$  into subfiles as
\begin{align}
W_{i}=\{W_{i,\Tc}:\Tc \subseteq [\Ksf]\}, \label{eq:subfile division}
\end{align}
where $W_{i,\Tc}$ represents the bits of $W_{i}$  exclusively cached by the cache nodes in  $\Tc$.
The optimal load  under the constraint of uncoded cache placement is denoted by $\Rsf^{\star}_{\rm u} $.

\subsection{Optimality of the MAN Scheme under Uncoded Cache Placement for the Shared-link Model}
\label{sub:shared-link}
In the following, we briefly introduce the MAN coded caching scheme~\cite{dvbt2fundamental} for the shared-link MAN coded caching model,  including  a server with $\Nsf $ files and $\Ksf $ cache-aided users with cache size $\Msf $.

We focus on the memory size $\Msf=\frac{\Nsf t}{\Ksf}$, where $t\in [0:\Ksf]$.
By dividing each file $W_i$ where $i\in [\Nsf]$ into ${\Ksf\choose t}$ non-overlapping and equal-length subfiles, $W_{i}=\{W_{i,\Tc}:\Tc\subseteq [\Ksf], |\Tc|=t\}$, we let each user $k\in [\Ksf]$ cache $W_{i,\Tc}$ where $k\in \Tc$.
Hence, each user totally caches $\Nsf \frac{\binom{\Ksf-1}{t-1}}{\binom{\Ksf}{t}} \Bsf=\frac{\Nsf t}{\Ksf}B=\Msf \Bsf$ bits, satisfying the memory size constraint.

In the delivery phase, we assume that the demand vector is ${\bf d}=(d_1,\ldots,d_{\Ksf}) \in [\Nsf]^{\Ksf}$. For each set $\Sc \subseteq [\Ksf]$ where $|\Sc|=t+1$, the server broadcasts a multicast message $X_{\Sc}=  \underset{k\in \Sc}{\oplus } W_{d_k,\Sc\setminus \{k\}}.$ Each user  $k\in \Sc$ caches all subfiles but $W_{d_k,\Sc\setminus \{k\}}$ in $X_{\Sc}$. Since the server   transmits $\binom{\Ksf}{t+1}$ multicast messages, each of which contains $\Bsf/\binom{\Ksf}{t}$ bits, the  achieved memory-load tradeoff is
\begin{align}
(\Msf, \Rsf_{\text{MAN}} )= \left( \frac{\Nsf t}{\Ksf}, \frac{\Ksf-t}{t+1} \right), \ \forall t\in[0:\Ksf]. \label{eq:MAN tradeoff}
\end{align}

The lower convex envelope of the memory-load tradeoff points in~\eqref{eq:MAN tradeoff} was proved to be optimal under the constraint of uncoded cache placement and $\Nsf  \geq \Ksf $~\cite{indexcodingcaching2020,exactrateuncoded}.
 More precisely, for any coded caching scheme with uncoded cache placement ${\bf Z}$, we can divide each file into $2^{\Ksf}$ subfiles as in~\eqref{eq:subfile division}.
  We consider one demand vector ${\bf d}=(d_1,\ldots, d_{\Ksf})$ where $d_{i} \neq d_j$ if $i\neq j$, and one permutation  of $[\Ksf]$ denoted by ${\bf u}=(u_1,\ldots,u_{\Ksf})$.
We then construct a genie-aided super-user with cached content
\begin{align*}
&Z^{\prime}=\big(Z_{u_1}, Z_{u_2}\setminus (W_{d_{u_1}}\cup Z_{u_1}), \ldots, \\& Z_{u_{\Ksf}}\setminus (W_{d_{u_1}}\cup Z_{u_1}\cup W_{d_{u_2}} \cup Z_{u_2} \cup  \cdots \cup W_{d_{u_{\Ksf-1}}}\cup Z_{u_{\Ksf-1}})  \big),
\end{align*}
who is able to recover  $(W_{d_1},\ldots, W_{d_{\Ksf}})$ from $(X ,Z^{\prime})$. This is because, this super-user   first decodes $W_{d_{u_1}}$ from $(X,Z_{u_1})$, then decodes $W_{d_{u_2}}$ from $(X,Z_{u_1}, Z_{u_2}\setminus (W_{d_{u_1}}\cup Z_{u_1}) )$, and does the similar procedure iteratively  until decoding   $W_{d_{u_{\Ksf}}}$.
Hence, we have
\begin{subequations}
  \begin{align}
 &    H(W_{d_1},\ldots, W_{d_{\Ksf}}|Z^{\prime}) = H(W_{d_1},\ldots, W_{d_{\Ksf}}|Z^{\prime},X) \nonumber\\& + I(W_{d_1},\ldots, W_{d_{\Ksf}};X|Z^{\prime}) \\&= I(W_{d_1},\ldots, W_{d_{\Ksf}};X|Z^{\prime})  \leq  H(X|Z^{\prime})  \leq H(X) , \\
% H(W_{d_1},\ldots, W_{d_{\Ksf}})  \leq H(X, Z^{\prime}) \leq H(X) + H(Z^{\prime}),\\
   & \Longrightarrow  \Rsf   \geq \sum_{i \in [\Ksf]} \  \sum_{\Tc \subseteq [\Ksf] \setminus \{ u_1,\ldots, u_i\} }  \frac{|W_{d_{u_i},\Tc}|}{\Bsf}. \label{eq:acyclic lower bound}
  \end{align}
 \end{subequations}

By considering all demand vectors in  which users have distinct demands and all permutations of users, we sum all the inequalities in the form of~\eqref{eq:acyclic lower bound}.  Because of symmetry, for each   $t\in [0:\Ksf]$, on the right side of the sum of all these inequalities, the coefficients of the term $|W_{i,\Tc}|$
are the same, where $i\in [\Nsf]$ and $|\Tc|=t$. Note that in~\eqref{eq:acyclic lower bound}, there are $\binom{\Ksf}{t+1}$ terms with $|\Tc| = t$ whose coefficient is $1$. Hence,   the sum of all inequalities in the form of~\eqref{eq:acyclic lower bound} is
\begin{subequations}
\begin{align}
& \Rsf   \geq  \sum_{t\in [0:\Ksf]} \frac{\binom{\Ksf}{t+1}}{\Nsf \binom{\Ksf}{t}} x_t =  \sum_{t\in [0:\Ksf]} \frac{\Ksf-t}{\Nsf(t+1)} x_t, \label{eq:sum ineq}\\
&\text{ where } x_t:= \sum_{i\in [\Nsf]} \  \sum_{\Tc \subseteq [\Ksf]:|\Tc|=t} \frac{|W_{i,\Tc}|}{\Bsf} . \label{eq:def of xt}
\end{align}
  \end{subequations}
From the memory size constraint, it should satisfy
\begin{align}
\sum_{t\in [0:\Ksf]} t x_t \leq  \Ksf \Msf. \label{eq:cache size}
\end{align}
From the file size constraint, it should satisfy
\begin{align}
\sum_{t\in [0:\Ksf]}  x_t = \Nsf. \label{eq:file size}
\end{align}

%For each $q\in [0:\Ksf-1]$, we perform Fourier Motzkin elimination on $x_q$ and $x_{q+1}$.
Finally, by the Fourier-Motzkin elimination  on $x_q$ where $q\in [0:\Ksf]$, we obtain $\Rsf^{\star}_{\rm u}$ is lower bounded
 by the lower convex envelope of the memory-load points $ \left( \frac{\Nsf t}{\Ksf}, \frac{\Ksf-t}{t+1} \right)$, where $t\in [0:\Ksf]$,    coinciding with that of the MAN scheme.

\section{Main Results}
\label{sec:results}
For the considered coded caching  problem  for location-based content,   the $\Ksf$-user MAN scheme could be directly used, which achieves the memory-load tradeoff points in~\eqref{eq:MAN tradeoff}. However, in the following theorem, we show that this topology-agnostic scheme is strictly sub-optimal.
%In this section, we first list our main results on the   $(\Ksf,\asf,\bsf)$   coded caching  problem  for location-based content downloads, and then provide the achivability and converse proofs in the rest of this section.
\begin{thm}
\label{thm:main result}
For the   $(\Ksf,\asf,\bsf)$   coded caching  problem  for location-based content,
when $\bsf(\Ksf-1)<2\asf$, the optimal load under the constraint of uncoded cache placement is
\begin{align}
\Rsf^{\star}_{\rm u} = \begin{cases}  \Ksf-\frac{\Ksf+1}{2(\asf+\bsf)}\Msf , & \text{ if } \  0 \leq \Msf \leq \asf +\bsf ; \\  \frac{(\Ksf-1)(2\asf+\bsf)}{2\asf}-\frac{\Ksf-1}{2\asf}\Msf, & \text{ if } \  \asf+\bsf < \Msf \leq 2\asf+\bsf. \end{cases} \label{eq:first regime}
\end{align}
When  $\bsf(\Ksf-1)\geq 2\asf$,  the optimal load under the constraint of uncoded cache placement is
\begin{align}
\Rsf^{\star}_{\rm u} =  \Ksf-\frac{\Ksf}{2\asf+\bsf}\Msf. \label{eq:sec regime}
\end{align}
\hfill $\square$
\end{thm}
\begin{IEEEproof}
 {\it Achievability.} When $\bsf(\Ksf-1)<2\asf$, the optimal load under   uncoded cache placement in~\eqref{eq:first regime} is achieved by the memory sharing among the memory-load points $(0,\Ksf)$, $\left(\asf+\bsf, \frac{\Ksf-1}{2} \right)$, and $(2\asf+\bsf, 0)$. For $(0,\Ksf)$, we let the server directly transmit $W_{d_k}$ for each $k\in [\Ksf]$.  For $\left(\asf+\bsf, \frac{\Ksf-1}{2} \right)= \left( \frac{\Nsf}{\Ksf}, \frac{\Ksf-1}{2} \right)$, we directly use the MAN scheme with $t=1$ in~\eqref{eq:MAN tradeoff}. For $(2\asf+\bsf, 0)$, we let each cache node $k$ cache  all the  $2\asf+\bsf$ files in $\Dc_k$. Since the demand of user  $k$ is in $\Dc_k$ and user $k$ can retrieve the cached content of cache node $k$, the load in the delivery phase is $0$.
  
When $\bsf(\Ksf-1)\geq 2\asf$, the optimal load under   uncoded cache placement in~\eqref{eq:sec regime} is achieved by the memory sharing between the memory-load points $(0,\Ksf)$  and $(2\asf+\bsf, 0)$, which can be achieved as described above.
 Note that  when $\bsf(\Ksf-1)> 2\asf$,  the MAN scheme is strictly sub-optimal for any    $0<  \Msf \leq 2\asf+\bsf$. 

 {\it Converse.}
The main technical challenge for Theorem~\ref{thm:main result} is the proof of the converse under uncoded cache placement. Since the set of possible demanded files by each user is a proper subset of $[\Nsf]$, the converse for the original MAN coded caching problem is not a converse for our considered problem. Similar to the converse bound for~\eqref{eq:MAN tradeoff}, we can consider all possible demand vectors where users have distinct demands and all permutations of the $\Ksf$ users, and obtain a lower bound on the load in the form of~\eqref{eq:acyclic lower bound} for each combination of the aforementioned demand vector and permutation. Together with the memory size and file size constraints, we can obtain a converse bound on $\Rsf^{\star}_{\rm u} $, which is a Linear Programming (LP) with the numbers of constraints and of variables exponential to $\Ksf$.  To compute the closed-form of the optimal solution for the LP, one idea is to sum all the inequalities in the form of~\eqref{eq:acyclic lower bound}  as we did for the original MAN coded caching problem.
However, summing all the inequalities loosens the converse bound in our problem, because some inequalities are redundant.\footnote{\label{foot:redundancy}For example, if we have two lower bounds on $\Rsf$, say  $\Rsf \geq 3$ and $\Rsf \geq 1$. Obviously, $\Rsf \geq 1$ is redundant. If we sum these two bounds, we have $\Rsf \geq 2$, which is   looser than   $\Rsf \geq 3$.}  Intuitively, this redundancy is  because in this network topology the demand vectors are not symmetric, neither the permutations of users; thus the resulting inequalities are not symmetric. {\bf Instead, our main contribution is to smartly select the non-redundant inequalities. This is done by 
carefully selecting the demand vectors and     specific permutation(s) of users  for each selected demand vector.} 
%This  is  different from the converse bound in Section~\ref{sub:shared-link} which considers all distinct demands and all permutations of users.
Then we  sum these non-redundant inequalities all together, such that we can obtain  a closed-form of the solution for the LP, which is exactly identical to the optimal load in Theorem~\ref{thm:main result}. The detailed proof on the converse bound for Theorem~\ref{thm:main result} could be found in Sections~\ref{sec:converse first seg} to~\ref{sec:converse thrid seg}.
\end{IEEEproof}

\begin{rem}[Effect of $\asf$ and $\bsf$]
\label{rem:discussion on main theorm}
 It is interesting to see  from Theorem~\ref{thm:main result} that, when   $\bsf\geq \frac{2\asf}{\Ksf-1}$, under the constraint of uncoded cache placement, coded caching   does not have any advantage compared to the uncoded caching  scheme 
 (i.e., the memory sharing between $(0,\Ksf)$  and $(2\asf+\bsf, 0)$). By contrast, when $\asf$ increases,  coded caching becomes more significant compared to uncoded caching, as illustrated in Fig.~\ref{fig:numerical}.   The coded caching gain of the proposed scheme  for Theorem~\ref{thm:main result} is no more than $2$ compared to the uncoded caching scheme, since we only use the MAN  coded caching scheme in~\eqref{eq:MAN tradeoff} with $t=1$. 
   \begin{figure}
    \centering
    \begin{subfigure}[t]{0.5\textwidth}
        \centering
        \includegraphics[scale=0.5]{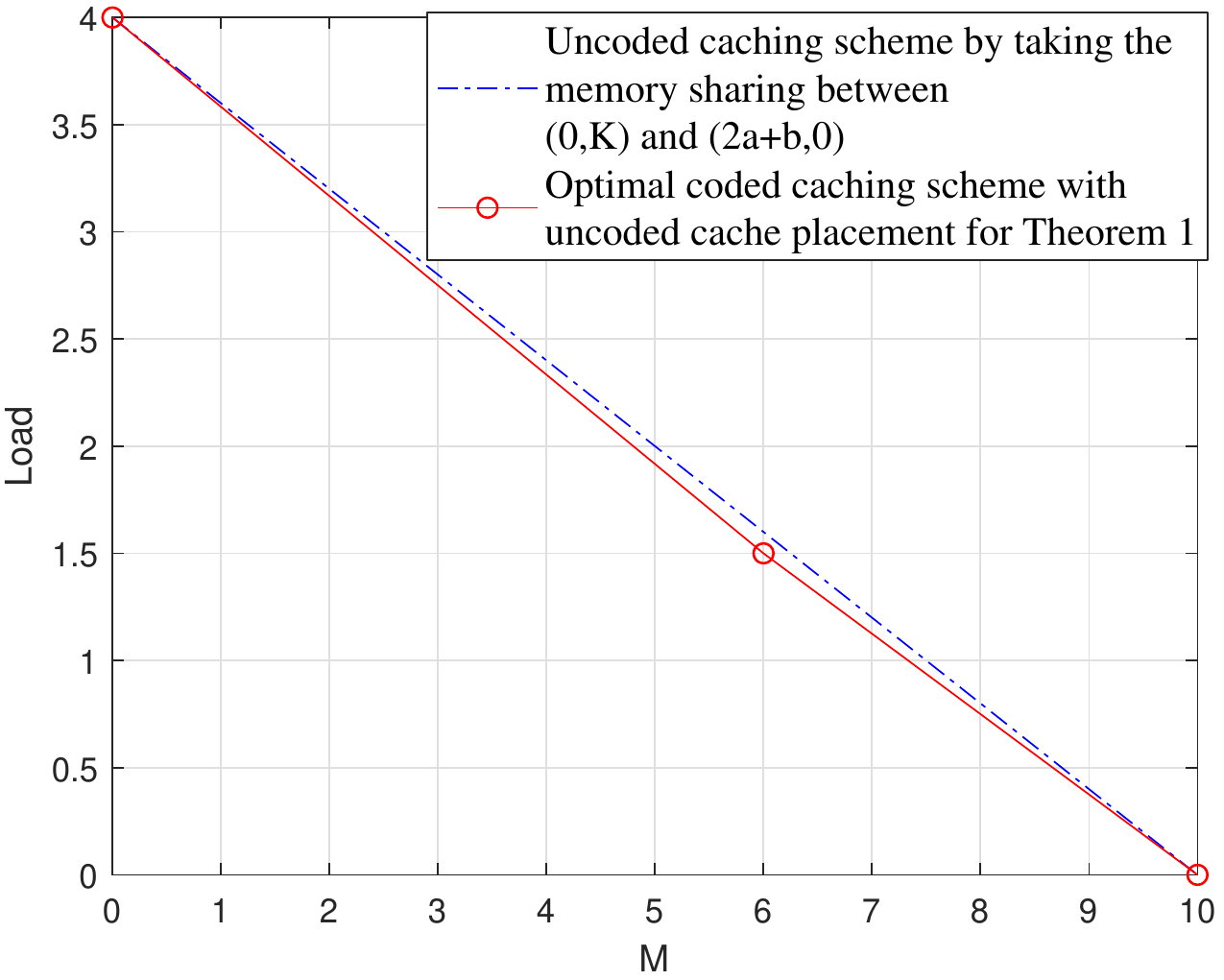}
        \caption{\small $\asf=4$.}
        \label{fig:numericala}
    \end{subfigure}%
    \\ 
    \begin{subfigure}[t]{0.5\textwidth}
        \centering
        \includegraphics[scale=0.5]{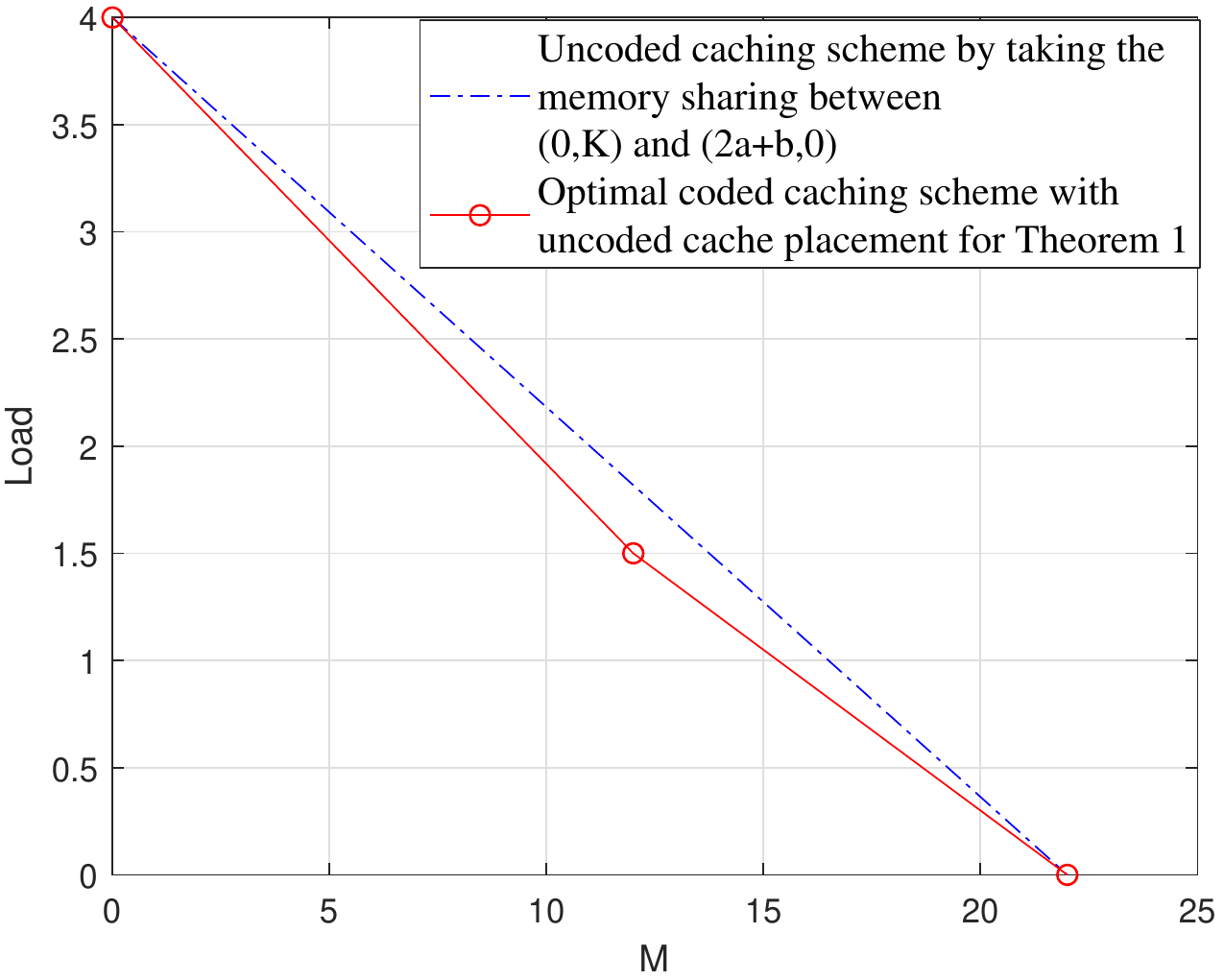}
        \caption{\small $\asf=10$.}
        \label{fig:numericalb}
    \end{subfigure}
        \caption{\small Coded caching  problem  for location-based content  with $\Ksf=4$,    $ \bsf=2$, and   various values  $\asf$.}
    \label{fig:numerical}
\end{figure}  
\hfill $\square$
\end{rem}

By comparing the achieved load in Theorem~\ref{thm:main result} with a   cut-set converse bound, we obtain the following order optimality results, whose proof could be found in Appendix~\ref{sec:converse order optimal}.
\begin{thm}
\label{thm:order optimal}
For the   $(\Ksf,\asf,\bsf)$   coded caching  problem  for location-based content,
\begin{itemize}
\item if $\Ksf$ is even, we have  $\Rsf^{\star}_{\rm u} \leq 2 \Rsf^{\star}$;
\item if $\Ksf$ is odd, we have  $\Rsf^{\star}_{\rm u} \leq 3 \Rsf^{\star}$.
\end{itemize}
\hfill $\square$
\end{thm}
Theorem~\ref{thm:order optimal} shows that the proposed achievable scheme  for  Theorem~\ref{thm:main result} is generally order optimal within a factor of $3$.

\begin{rem}[Extension to the multiple-input single-output (MISO)  broadcast channel]
\label{rem:extension to MISO BC}
The proposed achievable scheme in this paper 
 could be directly extended to the case where the server has multiple antennas by using the cache-aided MISO schemes in~\cite{interferencemanagement,degreesHachem2018,multiant2017sharia,addinganntenna2018Lamp}. By leveraging the multiplexing gain from $\Lsf$ antennas at the server,
 the achieved corner points become 
   $\left(0, \frac{\Ksf}{\Lsf} \right)$, $\left( \asf+\bsf, \frac{\Ksf-1}{\Lsf+1} \right)$, and $(2\asf+\bsf, 0)$.
\hfill $\square$
\end{rem}

\subsection{Converse Proof of Theorem~\ref{thm:main result}: $\bsf(\Ksf-1)<2\asf$ and $\asf+\bsf  \leq  \Msf \leq 2\asf+\bsf$}
\label{sec:converse first seg}
We first focus on the case where $\bsf(\Ksf-1)<2\asf$ and $\asf+\bsf  \leq  \Msf \leq 2\asf+\bsf$, and use the following example to illustrate the main idea of our proposed converse bound under uncoded cache placement.
\begin{example}[$(\Ksf,\asf,\bsf)=(3,2,1)$ and $3\leq \Msf \leq 5$]
\label{ex:case 1 first regime}
Consider the coded caching  problem  for location-based content   with $\Ksf=3$, $\asf=2$, and $\bsf=1$. In this example,   $\Nsf=\Ksf(\asf+\bsf)=9$ and
$$
\Dc_1= \{ 1,2,3,4,5\} , \ \Dc_2=\{4,5,6,7,8\}, \ \Dc_3=\{7,8,9,1,2\}.
$$
It can be seen that the $\Nsf=9$ files could be divided into two classes, where in the first class denoted by $\Cc_1=\{1,2,4,5,7,8 \}$, each file may be demanded by two users; in the second class denoted by $\Cc_2=\{3,6,9\}$, each file can only be demanded by one user.
The achieved load by the proposed scheme for Theorem~\ref{thm:main result} is $  \frac{5}{2}-\frac{\Msf}{2}$ when  $3\leq \Msf \leq 5$. In the following, we will prove that it is optimal under uncoded cache placement.

For any caching scheme with uncoded cache placement ${\bf Z}$, we can divide each file $W_i$,  $i\in [\Nsf]$, into subfiles $W_i=\{W_{i,\Tc}:\Tc\subseteq [3]\}$, where $W_{i,\Tc}$ represents the bits of $W_i$ exclusively cached by the cache nodes in $\Tc$.
Different from the converse proof for the MAN coded caching problem described in Section~\ref{sub:shared-link} which considers all possible demand vectors with
distinct demands and all permutations of $\Ksf$ users, we will carefully select the demand vectors and user permutations which lead to non-redundant inequalities on the load.

We first fix one permutation $(u_1,u_2,u_3)=(1,3,2)$. For this permutation of users,  we consider the demand vectors $(d_1,d_2,d_3)$  where     $d_1 \in \{1,2 \}$, $d_2=6$ and $d_3 \in\{7,8\}$. More precisely, pick one demand vector $(d_1,d_2,d_3)=(1,6,7)$, we construct a genie-aided super user with cache
\begin{align*}
Z^{\prime}&=\big(Z_{ u_1}, Z_{u_2}\setminus (Z_{u_1} \cup W_{d_{u_1}}), \\& Z_{u_3} \setminus (Z_{u_1}\cup W_{d_{u_1}} \cup Z_{u_2} \cup W_{d_{u_2}} ) \big) \\
& = \big(Z_{1}, Z_{3}\setminus (Z_{1} \cup W_{1}), Z_{2} \setminus (Z_{1}\cup W_{1} \cup Z_{3} \cup W_{7} ).
\end{align*}
From $(Z^{\prime},X)$, the virtual user can decode $W_{1}$, $W_{7}$, and $W_6$, iteratively. Hence, we have
 % \begin{subequations}
\begin{align}
&H(W_{1},W_{7},W_{6}|Z^{\prime})\nonumber\\&= H(W_{1},W_{7},W_{6}|Z^{\prime},X) + I(W_{1},W_{7},W_{6};X|Z^{\prime}) \nonumber\\& \leq H(X), \nonumber \\
 &  \Longrightarrow   \Rsf  \geq (|W_{1,\emptyset}| +|W_{1,\{2\}}| +  |W_{1,\{3\}}| +  |W_{1,\{2,3\}}| \nonumber\\& + |W_{7,\emptyset}| + |W_{7,\{2\}}| +|W_{6,\emptyset}|)/\Bsf, \nonumber\\
 &  \Longrightarrow \Rsf  \geq (|W_{1,\emptyset}| +|W_{1,\{2\}}| +  |W_{1,\{3\}}|    + |W_{7,\emptyset}| \nonumber\\& + |W_{7,\{2\}}| +|W_{6,\emptyset}|)/\Bsf. \label{eq:167 132}
\end{align}
%   \end{subequations}
Similarly, for this permutation of users,   when $(d_1,d_2,d_3)=(1,6,8)$, we have
\begin{align}
\Rsf & \geq (|W_{1,\emptyset}| +|W_{1,\{2\}}| +  |W_{1,\{3\}}|   + |W_{8,\emptyset}|  \nonumber\\& + |W_{8,\{2\}}| +|W_{6,\emptyset}|)/\Bsf. \label{eq:168 132}
\end{align}
When $(d_1,d_2,d_3)=(2,6,7)$, we have
\begin{align}
\Rsf  &\geq (|W_{2,\emptyset}| +|W_{2,\{2\}}| +  |W_{2,\{3\}}|   + |W_{7,\emptyset}| \nonumber\\& + |W_{7,\{2\}}| +|W_{6,\emptyset}|)/\Bsf. \label{eq:267 132}
\end{align}
When $(d_1,d_2,d_3)=(2,6,8)$, we have
\begin{align}
\Rsf  &\geq (|W_{2,\emptyset}| +|W_{2,\{2\}}| +  |W_{2,\{3\}}|   + |W_{8,\emptyset}| \nonumber\\& + |W_{8,\{2\}}| +|W_{6,\emptyset}|)/\Bsf. \label{eq:268 132}
\end{align}

We then fix one permutation $(u_1,u_2,u_3)=(1,2,3)$. For this permutation of users,  we consider the demand vectors $(d_1,d_2,d_3)$  where     $d_1 \in \{4,5 \}$,   $d_2\in \{7,8 \}$, and $d_3=9$.   More precisely, when $(d_1,d_2,d_3)=(4,7,9)$, we have
 \begin{align}
 \Rsf  &\geq (|W_{4,\emptyset}| +|W_{4,\{2\}}| +  |W_{4,\{3\}}|   + |W_{7,\emptyset}| \nonumber\\&+ |W_{7,\{3\}}| +|W_{9,\emptyset}|)/\Bsf. \label{eq:479 123}
 \end{align}
When $(d_1,d_2,d_3)=(4,8,9)$, we have
 \begin{align}
 \Rsf  &\geq (|W_{4,\emptyset}| +|W_{4,\{2\}}| +  |W_{4,\{3\}}|   + |W_{8,\emptyset}| \nonumber\\&+ |W_{8,\{3\}}| +|W_{9,\emptyset}|)/\Bsf. \label{eq:489 123}
 \end{align}
 When $(d_1,d_2,d_3)=(5,7,9)$, we have
 \begin{align}
 \Rsf  &\geq (|W_{5,\emptyset}| +|W_{5,\{2\}}| +  |W_{5,\{3\}}|   + |W_{7,\emptyset}| \nonumber\\&+ |W_{7,\{3\}}| +|W_{9,\emptyset}|)/\Bsf. \label{eq:579 123}
 \end{align}
  When $(d_1,d_2,d_3)=(5,8,9)$, we have
 \begin{align}
 \Rsf  &\geq (|W_{5,\emptyset}| +|W_{5,\{2\}}| +  |W_{5,\{3\}}|   + |W_{8,\emptyset}| \nonumber\\&+ |W_{8,\{3\}}| +|W_{9,\emptyset}|)/\Bsf. \label{eq:589 123}
 \end{align}

 By summing~\eqref{eq:167 132}-\eqref{eq:589 123},  we obtain
 \begin{align}
 \Rsf &\geq  \frac{1}{4 \Bsf}(|W_{1,\emptyset}|  + |W_{2,\emptyset}|+  |W_{4,\emptyset}| +|W_{5,\emptyset}| +2 |W_{7,\emptyset}| \nonumber\\& + 2|W_{8,\emptyset}| )  + \frac{1}{2\Bsf} (|W_{6,\emptyset}|+|W_{9,\emptyset}|)  \nonumber \\
 & + \frac{1}{4\Bsf} \sum_{i\in \{1,2,4,5,7,8\}} \sum_{j\in \{2,3\}} |W_{i,\{j\}}|.  \label{eq:ex1 permutation 1}
 \end{align}

 Next we fix one permutation $(u_1,u_2,u_3)=(2,1,3)$. For this permutation of users,  we consider the demand vectors $(d_1,d_2,d_3)$  where     $d_1\in \{1,2 \}$,   $d_2\in \{4,5 \}$, and $d_3=9$.
Hence, we can list $4$ inequalities.
 We also fix one permutation $(u_1,u_2,u_3)=(2,3,1)$.
 For this permutation of users,  we consider the demand vectors $(d_1,d_2,d_3)$  where    $d_1 =3$,   $d_2\in \{7,8 \}$, and $d_3\in \{ 1,2 \}$. Hence, we can also list $4$ inequalities.
Then we sum these $8$ inequalities  to obtain
\begin{align}
 \Rsf &\geq  \frac{1}{4 \Bsf}(|W_{4,\emptyset}|  + |W_{5,\emptyset}|+  |W_{7,\emptyset}| +|W_{8,\emptyset}| +2 |W_{1,\emptyset}| \nonumber \\
 &+ 2|W_{2,\emptyset}| )  + \frac{1}{2\Bsf} (|W_{9,\emptyset}|+|W_{3,\emptyset}|)  \nonumber \\
 &+  \frac{1}{4\Bsf} \sum_{i\in \{1,2,4,5,7,8\}} \sum_{j\in \{1,3\}} |W_{i,\{j\}}| . \label{eq:ex1 permutation 2}
\end{align}

Finally, we fix one permutation $(u_1,u_2,u_3)=(3,2,1)$. For this permutation of users, we consider the demand vectors $(d_1,d_2,d_3)$  where     $d_1 =3$,   $d_2\in \{4,5 \}$, and $d_3 \in\{7,8\}$. Hence, we can list $4$ inequalities.
 We also fix one permutation $(u_1,u_2,u_3)=(3,1,2)$. For this permutation of users,  we consider the demand vectors $(d_1,d_2,d_3)$  where    $d_1 \in\{ 4,5\}$,   $d_2=6$, and $d_3\in \{ 1,2 \}$. Hence, we can also list $4$ inequalities.
Then we sum these $8$ inequalities  to obtain
 \begin{align}
 \Rsf &\geq  \frac{1}{4 \Bsf}(|W_{7,\emptyset}|  + |W_{8,\emptyset}|+  |W_{1,\emptyset}| +|W_{2,\emptyset}| +2 |W_{4,\emptyset}|\nonumber \\
 & + 2|W_{5,\emptyset}| )  + \frac{1}{2\Bsf} (|W_{3,\emptyset}|+|W_{6,\emptyset}|)  \nonumber \\
&+  \frac{1}{4\Bsf} \sum_{i\in \{1,2,4,5,7,8\}} \sum_{j\in \{1,2\}} |W_{i,\{j\}}| . \label{eq:ex1 permutation 3}
\end{align}

By summing~\eqref{eq:ex1 permutation 1}-\eqref{eq:ex1 permutation 3}, we have
\begin{subequations}
\begin{align}
&\Rsf \geq\nonumber \\
 & \frac{1}{3}  \underbrace{(|W_{1,\emptyset}| +|W_{2,\emptyset}| + |W_{4,\emptyset}|  + |W_{5,\emptyset}| + |W_{7,\emptyset}|  + |W_{8,\emptyset}| )/\Bsf  }_{:=\alpha_0 } \nonumber \\
 & + \frac{1}{3}  \underbrace{(|W_{3,\emptyset}| +|W_{6,\emptyset}| + |W_{9,\emptyset}|) /\Bsf  }_{:=\beta_0 } \nonumber\\
& +\frac{1}{6}   \underbrace{ \sum_{i\in \{1,2,4,5,7,8\}} \sum_{j\in [3]} |W_{i,\{j\}}|/\Bsf }_{:=\alpha_1 }  \label{eq:ex1 final sum}\\
&= \frac{1}{3} \alpha_0 + \frac{1}{3} \beta_0 + \frac{1}{6} \alpha_1.
\label{eq:ex1 final sum ab}
\end{align}
\end{subequations}

By the file size constraint, for the first class of files $\Cc_1$, we have
\begin{subequations}
\begin{align}
6&=(|W_{1}|+|W_{2}|+|W_{4}|+|W_{5}|+|W_{7}|+|W_{8}|)/\Bsf  \\
&= \alpha_0 +\alpha_1+    \sum_{t_1 \in \{2,3\}} \sum_{i_1 \in \{1,2,4,5,7,8 \} } \sum_{\Tc_1\subseteq [3]:|\Tc_1|=t_1 } \negmedspace\negmedspace \frac{|W_{i_1,\Tc_1}|}{\Bsf}; \label{eq:ex1 file 1}
\end{align}
\end{subequations}
and  for the second class of files $\Cc_2$, we have
\begin{subequations}
\begin{align}
3&=(|W_{3}|+|W_{6}|+|W_{9}| )/\Bsf   \\& = \beta_0 +  \sum_{t_2\in [3]} \sum_{i_2\in \{3,6,9 \} } \sum_{\Tc_2\subseteq [3]:|\Tc_2|=t_2} \frac{|W_{i_2,\Tc_2}|}{\Bsf}. \label{eq:ex1 file 2}
\end{align}
\end{subequations}

 By the memory size constraint, we have
 \begin{align}
3\Msf &\geq  \alpha_1 +  \sum_{t_1 \in \{2,3\}} \sum_{i_1\in \{1,2,4,5,7,8 \} } \sum_{\Tc_1\subseteq [3]:|\Tc_1|=t_1 } \negmedspace\negmedspace \negmedspace\negmedspace \frac{t_1  |W_{i_1,\Tc_1}|}{\Bsf}  \nonumber\\& +
 \sum_{t_2\in [3]} \sum_{i_2\in \{3,6,9 \} } \sum_{\Tc_2\subseteq [3]:|\Tc_2|=t_2} \negmedspace\negmedspace \negmedspace\negmedspace  \frac{t_2|W_{i_2,\Tc_2}|}{\Bsf} .\label{eq:ex1 memory size}
 \end{align}

The next step is to derive the converse bound on $\Rsf$ from the constraints in~\eqref{eq:ex1 final sum ab},~\eqref{eq:ex1 file 1},~\eqref{eq:ex1 file 2}, and~\eqref{eq:ex1 memory size}.
More precisely, from~\eqref{eq:ex1 file 1} we have
\begin{align}
&\frac{1}{3}(\alpha_0 +\alpha_1 ) + \frac{1}{3}\sum_{t_1 \in \{2,3\}} \sum_{i_1\in \{1,2,4,5,7,8 \} } \sum_{\Tc_1\subseteq [3]:|\Tc_1|=t_1 } \frac{|W_{i_1,\Tc_1}|}{\Bsf}\nonumber\\& =2.\label{eq:ex1 cons1}
\end{align}
From~\eqref{eq:ex1 file 2} we have
\begin{align}
\frac{1}{6}\beta_0 +  \frac{1}{6}\sum_{t_2\in [3]} \sum_{i_2\in \{3,6,9 \} } \sum_{\Tc_2\subseteq [3]:|\Tc_2|=t_2} \frac{|W_{i_2,\Tc_2}|}{\Bsf}=\frac{1}{2}.\label{eq:ex1 cons2}
\end{align}
From~\eqref{eq:ex1 memory size} we have
\begin{align}
&-\frac{1}{6} \alpha_1  -\frac{1}{6}  \sum_{t_1 \in \{2,3\}} \sum_{i_1\in \{1,2,4,5,7,8 \} } \sum_{\Tc_1\subseteq [3]:|\Tc_1|=t_1 } \frac{t_1  |W_{i_1,\Tc_1}|}{\Bsf} \nonumber\\& -\frac{1}{6}
 \sum_{t_2\in [3]} \sum_{i_2\in \{3,6,9 \} } \sum_{\Tc_2\subseteq [3]:|\Tc_2|=t_2} \frac{t_2|W_{i_2,\Tc_2}|}{\Bsf} \geq -\frac{\Msf}{2}.\label{eq:ex1 cons3}
\end{align}
We sum~\eqref{eq:ex1 cons1}-\eqref{eq:ex1 cons3} to obtain
\begin{subequations}
\begin{align}
&\frac{1}{3} \alpha_0 + \frac{1}{6}\beta_0 + \frac{1}{6} \alpha_1  
  \geq \frac{5}{2}-\frac{\Msf}{2}  \nonumber\\&+ 
\frac{1}{6}  \sum_{t_1 \in \{2,3\}} \sum_{i_1\in \{1,2,4,5,7,8 \} } \sum_{\Tc_1\subseteq [3]:|\Tc_1|=t_1 } \left(   t_1 -2 \right) \frac{  |W_{i_1,\Tc_1}|}{\Bsf} \nonumber\\ & + 
\frac{1}{6} \sum_{t_2\in [3]} \sum_{i_2\in \{3,6,9 \} } \sum_{\Tc_2\subseteq [3]:|\Tc_2|=t_2} \left( t_2 -1 \right) \frac{|W_{i_2,\Tc_2}|}{\Bsf}   \\
& \geq \frac{5}{2}-\frac{\Msf}{2}
.\label{eq:ex1 before final}
\end{align}
\end{subequations}
By taking~\eqref{eq:ex1 before final} into~\eqref{eq:ex1 final sum ab}, we have
\begin{align}
\Rsf\geq \frac{1}{3} \alpha_0 + \frac{1}{3} \beta_0 + \frac{1}{6} \alpha_1   \geq \frac{1}{3} \alpha_0 + \frac{1}{6}\beta_0 + \frac{1}{6} \alpha_1  \geq   \frac{5}{2}-\frac{\Msf}{2}.\label{eq:ex1 final converse}
\end{align}
Hence, from~\eqref{eq:ex1 final converse} we have $\Rsf^{\star}_{\rm u} \geq   \frac{5}{2}-\frac{\Msf}{2}$,
which coincides with the achieved load for Theorem~\ref{thm:main result} when $3\leq \Msf \leq 5$.

Note that  if we consider all the possible demand vectors with distinct demands and all permutations of users, and sum all the obtained inequalities  from them, the resulting converse bound is not tight;  for example, if $\Msf=3$, the resulting non-tight converse bound provides $\Rsf^{\star}_{\rm u} \geq  \frac{54}{95}$, while the tight converse bound is $\Rsf^{\star}_{\rm u} \geq   \frac{5}{2}-\frac{\Msf}{2}=1$.
\hfill $\square$
\end{example}

We now generalize the converse bound proof in Example~\ref{ex:case 1 first regime} for the case $\bsf(\Ksf-1)<2\asf$ and $\asf+\bsf  \leq  \Msf \leq 2\asf+\bsf$. Recall that for each user $k\in [\Ksf]$, the set of possible demanded files by user $k$ is $\Dc_k$ defined in~\eqref{eq:Dk}, where $ \Dc_k:= \Dc_{k,1} \cup \Dc_{k,2} \cup \Dc_{k,3}$.
\iffalse
\begin{align}
 \Dc_k&:=  \underbrace{ \left[(k-1)(\asf+\bsf)+1 :  k\asf+(k-1)\bsf\right] }_{:=\Dc_{k,1} }  \cup  \underbrace{ \left[k\asf+(k-1)\bsf+1 :  k(\asf+\bsf)\right]  }_{:=\Dc_{k,2} }  \nonumber\\& \cup   \underbrace{  \left[< k(\asf+\bsf)+1 >_{  \Ksf(\asf+\bsf) }:<(k+1)\asf+ k \bsf >_{  \Ksf(\asf+\bsf) } \right]   }_{:=\Dc_{k,3} }.
\end{align}
\fi 
By definition,  $\Dc_{k,1}= \Dc_{<k-1>_{\Ksf},3}$ and $\Dc_{k,3}= \Dc_{<k+1>_{\Ksf},1}$. We also have $|\Dc_{k,1}|=|\Dc_{k,3}|=\asf$ and $|\Dc_{k,2}|=\bsf$.
 In addition, as in Example~\ref{ex:case 1 first regime}, we divide all the $\Nsf:=\Ksf(\asf+\bsf)$ files into two classes, where
 \begin{subequations}
 \begin{align}
& \Cc_1 := \cup_{k\in [\Ksf]} \Dc_{k,1} , \label{eq:def of C1}\\
&\text{and } \Cc_2 := \cup_{k\in [\Ksf]} \Dc_{k,2} . \label{eq:def of C2}
 \end{align}
\end{subequations}
For any caching scheme with uncoded cache placement ${\bf Z}$, we can divide each file $W_i$,  $i\in [\Nsf]$, into subfiles $W_i=\{W_{i,\Tc}:\Tc\subseteq [\Ksf]\}$.

Fix one integer $k\in [\Ksf]$. For this integer, we consider two permutations of users, $(k, <k-1>_{\Ksf},\ldots, <k-\Ksf+1>_{\Ksf})$ and
$(k, <k+1>_{\Ksf},\ldots, <k+\Ksf-1>_{\Ksf})$.

For the first permutation $(u_1,u_2,\ldots,u_{\Ksf})=(k, <k-1>_{\Ksf},\ldots, <k-\Ksf+1>_{\Ksf})$, we    consider the demand vectors $(d_1,\ldots,d_{\Ksf})$ where
$d_{u_j}\in\Dc_{u_j,1}$ for $j\in [\Ksf-1]$, and $d_{u_{\Ksf}}\in \Dc_{u_{\Ksf},2}$, totally $\asf^{\Ksf-1}\bsf$ demand vectors.
 For each $(d_1,\ldots,d_{\Ksf})$, we construct a genie-aided super user with cache
\begin{align}
Z^{\prime}&= \big( Z_{u_1},  Z_{u_2} \setminus (W_{d_{u_1}}\cup Z_{u_1}),   \ldots,   \nonumber\\&  Z_{u_{\Ksf}}\setminus ( W_{d_{u_1}}\cup Z_{u_1} \cup \cdots \cup W_{d_{u_{\Ksf-1}}}\cup Z_{u_{\Ksf-1}})    \big)  . \label{eq:cache vitrual user}
\end{align}
From $(X, Z^{\prime})$ we can decode $W_{d_{u_1}},\ldots, W_{u_{\Ksf}}$, iteratively. Hence,
 \begin{subequations}
\begin{align}
&  H(W_{d_{u_1}},\ldots, W_{d_{u_{\Ksf}}}|Z^{\prime})= H(W_{d_{u_1}},\ldots, W_{d_{u_{\Ksf}}}|Z^{\prime},X)\nonumber\\& + I(W_{d_{u_1}},\ldots, W_{d_{u_{\Ksf}}};X|Z^{\prime}) \\&=  I(W_{d_{u_1}},\ldots, W_{d_{u_{\Ksf}}};X|Z^{\prime}) \leq H(X),
\end{align}
 \end{subequations}
which leads to
 \begin{subequations}
\begin{align}
& \Rsf  \geq   \left(|W_{d_{u_1},\emptyset}|+\sum_{j_1\in [\Ksf]\setminus\{u_1\}} |W_{d_{u_1}, \{j_1\}}|\right)/\Bsf \nonumber\\& +
\left(|W_{d_{u_2},\emptyset}|+\sum_{j_2\in [\Ksf]\setminus\{u_1,u_2\}} |W_{d_{u_2}, \{j_2\}}|\right)/\Bsf \nonumber\\
& +\cdots+   |W_{d_{u_{\Ksf}},\emptyset}| /\Bsf \label{eq:general form with u of acyclic first perm}  \\
& =\left(|W_{d_{k},\emptyset}|+\sum_{j_1\in [\Ksf]\setminus\{k\}} |W_{d_{k}, \{j_1\}}|\right)/\Bsf \nonumber\\& +
\left(|W_{d_{<k-1>_{\Ksf}},\emptyset}|+\sum_{j_2\in [\Ksf]\setminus\{k,<k-1>_{\Ksf}\}} |W_{d_{<k-1>_{\Ksf}}, \{j_2\}}|\right)/\Bsf \nonumber\\&+\cdots+   |W_{d_{<k-\Ksf+1>_{\Ksf}},\emptyset}| /\Bsf.
\label{eq:general form of acyclic first perm}
\end{align}
  \end{subequations}
 Considering all  the demand vectors $(d_1,\ldots,d_{\Ksf})$ where
$d_{u_j}\in\Dc_{u_j,1}$ for $j\in [\Ksf-1]$, and $d_{u_{\Ksf}}\in \Dc_{u_{\Ksf},2}$, we list  $\asf^{\Ksf-1}\bsf$ inequalities in the form of~\eqref{eq:general form of acyclic first perm} and sum them all together to  obtain
\begin{align}
%\begin{split}
&\Rsf \geq  \frac{1}{\asf\Bsf} \sum_{i_1\in \Dc_{k,1}} \left(|W_{i_1,\emptyset}|+\sum_{j_1\in [\Ksf]\setminus\{k\}} |W_{i_1, \{j_1\}}|  \right) \nonumber\\& +   \frac{1}{\asf\Bsf} \sum_{i_2\in \Dc_{<k-1>_{\Ksf},1}} \left(|W_{i_2,\emptyset}|+\negmedspace\negmedspace\negmedspace\negmedspace \negmedspace\negmedspace\negmedspace\negmedspace \sum_{j_2\in [\Ksf]\setminus\{k, <k-1>_{\Ksf}\}} \negmedspace\negmedspace\negmedspace\negmedspace \negmedspace\negmedspace\negmedspace\negmedspace |W_{i_2, \{j_2\}}|  \right) + \cdots +\nonumber \\
&  \frac{1}{\asf \Bsf} \sum_{i_{\Ksf-1}\in \Dc_{<k-\Ksf+2>_{\Ksf},1}} \Bigg(|W_{i_{\Ksf-1},\emptyset}|   +\negmedspace\negmedspace\negmedspace\negmedspace \negmedspace \sum_{ \substack{ j_{\Ksf-1}\in [\Ksf]\setminus  \{k,\\ <k-1>_{\Ksf},\ldots, <k-\Ksf+2>_{\Ksf} \}}} \negmedspace\negmedspace\negmedspace\negmedspace\negmedspace\negmedspace\negmedspace \negmedspace \nonumber\\&  |W_{i_{\Ksf-1}, \{j_{\Ksf-1}\}}|  \Bigg) +    \frac{1}{\bsf \Bsf} \sum_{i_{\Ksf}\in \Dc_{<k-\Ksf+1>_{\Ksf},2}} |W_{i_{\Ksf},\emptyset}|.
\label{eq:general sum form of acyclic first perm}
%\end{split}
\end{align}

For the second permutation $(u_1,u_2,\ldots,u_{\Ksf})=(k, <k+1>_{\Ksf},\ldots, <k+\Ksf-1>_{\Ksf})$, we    consider the demand vectors $(d_1,\ldots,d_{\Ksf})$ where
$d_{u_j}\in\Dc_{u_j,3}$ for $j\in [\Ksf-1]$, and $d_{u_{\Ksf}}\in \Dc_{u_{\Ksf},2}$, totally $\asf^{\Ksf-1}\bsf$ demand vectors.
For each of such demand vectors, we construct a genie-aided super user with cache as in~\eqref{eq:cache vitrual user} and obtain an inequality as in~\eqref{eq:general form with u of acyclic first perm}. By summing all the obtained $\asf^{\Ksf-1}\bsf$ inequalities, we have
 \begin{subequations}
\begin{align}
& \Rsf \geq  \frac{1}{\asf\Bsf} \sum_{i_1\in \Dc_{k,3}} \left(|W_{i_1,\emptyset}|+\sum_{j_1\in [\Ksf]\setminus\{k\}} |W_{i_1, \{j_1\}}|  \right)  \nonumber\\& +
\frac{1}{\asf\Bsf} \sum_{i_2\in \Dc_{<k+1>_{\Ksf},3}} \left(|W_{i_2,\emptyset}|+ \negmedspace\negmedspace\negmedspace\negmedspace  \sum_{j_2\in [\Ksf]\setminus\{k,<k+1>_{\Ksf}\}}   \negmedspace\negmedspace\negmedspace\negmedspace\negmedspace\negmedspace \negmedspace\negmedspace\negmedspace  |W_{i_2, \{j_2\}}|  \right)
 \nonumber\\
& + \cdots +  \frac{1}{\asf \Bsf} \sum_{i_{\Ksf-1}\in \Dc_{<k+\Ksf-2>_{\Ksf},3}} \Bigg(|W_{i_{\Ksf-1},\emptyset}|  +\negmedspace\negmedspace\negmedspace\negmedspace \negmedspace\negmedspace\negmedspace\negmedspace \sum_{\substack{j_{\Ksf-1}\in [\Ksf]\setminus\{k,\\ <k+1>_{\Ksf},\ldots, <k+\Ksf-2>_{\Ksf}\}}} \nonumber\\&  |W_{i_{\Ksf-1}, \{j_{\Ksf-1}\}}|  \Bigg)
   + \frac{1}{\bsf \Bsf} \sum_{i_{\Ksf}\in \Dc_{<k+\Ksf-1>_{\Ksf},2}} |W_{i_{\Ksf},\emptyset}|   \\
&=  \frac{1}{\asf\Bsf}    \sum_{i_1\in \Dc_{<k+1>_{\Ksf} ,1}} \negmedspace\negmedspace\negmedspace  \left(|W_{i_1,\emptyset}|+\sum_{j_1\in [\Ksf]\setminus\{k\}} \negmedspace\negmedspace  |W_{i_1, \{j_1\}}|  \right)  \nonumber\\& +
\frac{1}{\asf\Bsf}    \sum_{i_2\in \Dc_{<k+2>_{\Ksf},3}} \left(|W_{i_2,\emptyset}|+ \negmedspace\negmedspace\negmedspace\negmedspace  \sum_{j_2\in [\Ksf]\setminus\{k,<k+1>_{\Ksf}\}}   \negmedspace\negmedspace\negmedspace\negmedspace\negmedspace\negmedspace \negmedspace\negmedspace\negmedspace  |W_{i_2, \{j_2\}}|  \right)
 \nonumber\\ & + \cdots + \frac{1}{\asf \Bsf} \negmedspace \sum_{i_{\Ksf-1}\in \Dc_{<k+\Ksf-1>_{\Ksf},1}}  \negmedspace\negmedspace\negmedspace \Bigg(|W_{i_{\Ksf-1},\emptyset}|+  \negmedspace\negmedspace\negmedspace\negmedspace\negmedspace\negmedspace\sum_{\substack{j_{\Ksf-1}\in [\Ksf]\setminus\{k, \\ <k+1>_{\Ksf}, \ldots, <k+\Ksf-2>_{\Ksf}\}}} \nonumber\\&  \negmedspace\negmedspace |W_{i_{\Ksf-1}, \{j_{\Ksf-1}\}}|  \Bigg)
  + \frac{1}{\bsf \Bsf} \sum_{i_{\Ksf}\in \Dc_{<k+\Ksf-1>_{\Ksf},2}} |W_{i_{\Ksf},\emptyset}|
  ,\label{eq:general sum form of acyclic sec perm}
\end{align}
  \end{subequations}
where~\eqref{eq:general sum form of acyclic sec perm} comes from that $\Dc_{i,3}= \Dc_{<i+1>_{\Ksf},1}$ for any $i\in [\Ksf]$.

We sum~\eqref{eq:general sum form of acyclic first perm}  and~\eqref{eq:general sum form of acyclic sec perm} to obtain
\begin{align}
\Rsf &\geq  \frac{1}{2\asf } \sum_{k_1\in \{k,<k+1>_{\Ksf}\}} \sum_{i_1\in \Dc_{k_1,1} }\frac{|W_{i_1,\emptyset}|}{\Bsf} \nonumber\\& + \frac{1}{\asf } \sum_{k_2 \in ([\Ksf]\setminus \{k,<k+1>_{\Ksf}\}) } \sum_{i_2\in \Dc_{k_2,1}}  \frac{|W_{i_2,\emptyset}|}{\Bsf}  \nonumber\\& + \frac{1}{2\bsf  } \sum_{k_3\in \{<k-1>_{\Ksf},<k+1>_{\Ksf}\}} \sum_{i_3\in \Dc_{k_3,2}  } \frac{|W_{i_3,\emptyset}|}{\Bsf}  \nonumber\\& +  \frac{1}{2\asf } \sum_{i_4\in \Cc_1} \sum_{j\in [\Ksf]\setminus\{k\}} \frac{|W_{i_4,\{j\}}|}{\Bsf},\label{eq:first regime begin with k}
\end{align}
where $\Cc_1=\cup_{k\in [\Ksf]} \Dc_{k,1}$ defined in~\eqref{eq:def of C1}.

By considering all $k\in [\Ksf]$, we list $\Ksf$ inequalities in the form of~\eqref{eq:first regime begin with k}, and then sum them all together to obtain
\begin{align}
\Rsf &\geq  \frac{\Ksf-1}{\asf \Ksf} \underbrace{ \sum_{i_1\in \Cc_1} \frac{|W_{i_1,\emptyset}|}{\Bsf}}_{:=\alpha_0 } +  \frac{1}{\bsf \Ksf} \underbrace{ \sum_{i_2\in \Cc_2} \frac{|W_{i_2,\emptyset}|}{\Bsf}  }_{:=\beta_0 }
\nonumber\\& + \frac{\Ksf-1}{2\asf \Ksf}  \underbrace{ \sum_{i_3\in \Cc_1} \sum_{j\in [\Ksf] } \frac{|W_{i_3,\{j\}}|}{\Bsf} }_{:=\alpha_1 }.
 \label{eq:first regime final sum}
 \end{align}

By the file size constraint, for the first class of files $\Cc_1$, we have
\begin{align}
 \alpha_0 +\alpha_1+    \sum_{t_1 \in [2:\Ksf]} \sum_{i_1\in \Cc_1 } \sum_{\Tc_1\subseteq [\Ksf]:|\Tc_1|=t_1 } \frac{|W_{i_1,\Tc_1}|}{\Bsf}= \asf \Ksf ; \label{eq:first regime file 1}
\end{align}
and  for the second class of files $\Cc_2$, we have
\begin{align}
  \beta_0 +  \sum_{t_2\in [\Ksf]} \sum_{i_2\in \Cc_2 } \sum_{\Tc_2\subseteq [\Ksf]:|\Tc_2|=t_2} \frac{|W_{i_2,\Tc_2}|}{\Bsf} =\bsf \Ksf. \label{eq:first regime file 2}
\end{align}

 By the memory size constraint, we have
 \begin{align}
  &\alpha_1 +  \sum_{t_1 \in [2:\Ksf]} \sum_{i_1\in \Cc_1 } \sum_{\Tc_1\subseteq [\Ksf]:|\Tc_1|=t_1 } \frac{t_1  |W_{i_1,\Tc_1}|}{\Bsf} \nonumber\\& +
 \sum_{t_2\in [\Ksf]} \sum_{i_2\in \Cc_2 } \sum_{\Tc_2\subseteq [\Ksf]:|\Tc_2|=t_2} \frac{t_2|W_{i_2,\Tc_2}|}{\Bsf} \leq \Ksf \Msf.\label{eq:first regime memory size}
 \end{align}

 We take $\frac{\Ksf-1}{\asf\Ksf} \times \eqref{eq:first regime file 1} + \frac{\Ksf-1}{2\asf \Ksf} \times \eqref{eq:first regime file 2}- \frac{\Ksf-1}{2\asf\Ksf} \eqref{eq:first regime memory size}$ to obtain
   \begin{subequations}
 \begin{align}
& \frac{\Ksf-1}{\asf\Ksf}  \alpha_0+ \frac{\Ksf-1}{2\asf \Ksf}   \beta_0  + \frac{\Ksf-1}{2\asf\Ksf}  \alpha_1  \nonumber\\
&\geq  \frac{(\Ksf-1)(2\asf+\bsf)}{2\asf}-\frac{\Ksf-1}{2\asf}\Msf
 \nonumber\\& +  \frac{\Ksf-1}{2\asf\Ksf} \sum_{t_1 \in [2:\Ksf]} \sum_{i_1\in \Cc_1 } \sum_{\Tc_1\subseteq [\Ksf]:|\Tc_1|=t_1 } ( t_1-2) \frac{|W_{i_1,\Tc_1}|}{\Bsf} \nonumber\\
 &+ \frac{\Ksf-1}{2\asf\Ksf}  \sum_{t_2\in [\Ksf]} \sum_{i_2\in \Cc_2 } \sum_{\Tc_2\subseteq [\Ksf]:|\Tc_2|=t_2} ( t_2-1) \frac{t_2|W_{i_2,\Tc_2}|}{\Bsf}  \\
 &\geq \frac{(\Ksf-1)(2\asf+\bsf)}{2\asf}-\frac{\Ksf-1}{2\asf}\Msf .
 \label{eq:first regime mf size}
 \end{align}
   \end{subequations}

By taking~\eqref{eq:first regime mf size} into~\eqref{eq:first regime final sum}, we have
   \begin{subequations}
\begin{align}
\Rsf &\geq   \frac{\Ksf-1}{\asf\Ksf}  \alpha_0+ \frac{1}{ \bsf \Ksf}   \beta_0  + \frac{\Ksf-1}{2\asf\Ksf}  \alpha_1  \\ & \geq  \frac{\Ksf-1}{\asf\Ksf}  \alpha_0+ \frac{\Ksf-1}{2\asf \Ksf}   \beta_0  + \frac{\Ksf-1}{2\asf\Ksf}  \alpha_1  \\
&\geq  \frac{(\Ksf-1)(2\asf+\bsf)}{2\asf}-\frac{\Ksf-1}{2\asf}\Msf ,  \label{eq:first regime final proof}
\end{align}
    \end{subequations}
    which leads to $\Rsf^{\star}_{\rm u} \geq  \frac{(\Ksf-1)(2\asf+\bsf)}{2\asf}-\frac{\Ksf-1}{2\asf}\Msf $,  coinciding with the achieved load for Theorem~\ref{thm:main result} when $\bsf(\Ksf-1)<2\asf$  and $\asf+\bsf \leq  \Msf \leq 2\asf+\bsf$.

    \subsection{Converse Proof of Theorem~\ref{thm:main result}: $\bsf(\Ksf-1)<2\asf$ and $0  \leq  \Msf \leq  \asf+\bsf$}
\label{sec:converse sec seg}
    We then focus on the case where $\bsf(\Ksf-1)<2\asf$ and $0  \leq  \Msf \leq \asf+\bsf$. We go back to Example~\ref{ex:case 1 first regime} and consider the  memory size regime $0\leq \Msf \leq 3$.
\begin{example}[$(\Ksf,\asf,\bsf)=(3,2,1)$ and $0\leq \Msf \leq 3$]
\label{ex:case 1 sec regime}
Recall that in this example we have   $\Nsf=\Ksf(\asf+\bsf)=9$ and
$ 
\Dc_1= \{ 1,2,3,4,5\}$ , $ \Dc_2=\{4,5,6,7,8\}$, $ \Dc_3=\{7,8,9,1,2\}.$ 
The achieved load by the proposed scheme for Theorem~\ref{thm:main result} is $  3-\frac{2}{3}\Msf$ when  $0\leq \Msf \leq 3$. In the following, we will prove that it is optimal under uncoded cache placement.

For any caching scheme with uncoded cache placement ${\bf Z}$,
with the definition of $\alpha_0$, $\beta_0$, and $\alpha_1$ given in~\eqref{eq:ex1 final sum}, it has been proved in~\eqref{eq:ex1 final sum ab} that
\begin{align}
\Rsf \geq \frac{1}{3} \alpha_0 + \frac{1}{3} \beta_0 + \frac{1}{6} \alpha_1. \label{eq:recal ex1 final sum ab}
\end{align}
We will derive another lower bound  for $\Rsf$  in terms of  $\alpha_0$, $\beta_0$, and $\alpha_1$, by using another strategy to select demand vectors and permutation of users.

We first fix one permutation $(u_1,u_2,u_3)=(1,3,2)$.    For this permutation of users,  we consider the demand vectors $(d_1,d_2,d_3)$  where     $d_1 \in \{1,2 \} $,   $d_2\in \{4,5 \}$, and $d_3 \in \{7,8\}$. For each of such $8$  demand vectors, we can generate an inequality on $\Rsf$; for example if $(d_1,d_2,d_3)=(1,4,7)$, by generating a genie-aided super user with cache  $Z^{\prime}=(Z_1, Z_3\setminus (W_1\cup Z_1), Z_2 \setminus (W_1\cup Z_1 \cup W_7 \cup Z_3))$, we can recover $(W_1,W_4,W_7)$ from $(Z^{\prime},X)$,  and thus
\begin{align}
\Rsf &\geq |W_{1,\emptyset}|/\Bsf +|W_{1, \{2\}}|/\Bsf  +|W_{1,\{3\}}| /\Bsf +|W_{7,\emptyset}|/\Bsf  \nonumber\\& +|W_{7,\{2\}}|/\Bsf  +|W_{4,\emptyset}|/\Bsf  . \label{eq:ex2 132 147}
\end{align}
By considering all such $8$  demand vectors, we can list $8$ inequalities in the form of~\eqref{eq:ex2 132 147}, and sum them all together to obtain
\begin{align}
\Rsf  &\geq  \frac{1}{2\Bsf}(|W_{1,\emptyset}|+ |W_{2,\emptyset}| + |W_{4,\emptyset}|+|W_{5,\emptyset}|+|W_{7,\emptyset}|\nonumber\\& +|W_{8,\emptyset}|) 
+ \frac{1}{2\Bsf} (|W_{1,\{2\}}| +|W_{1,\{3\}}| +|W_{2,\{2\}}|+|W_{2,\{3\}}| \nonumber\\& +|W_{7,\{2\}}|  +|W_{8,\{2\}}|    ). \label{eq:ex2 132}
\end{align}
 We then fix one permutation $(u_1,u_2,u_3)=(1,2,3)$.    For this permutation of users,  we consider the demand vectors $(d_1,d_2,d_3)$  where     $d_1 \in \{4,5 \} $,   $d_2\in \{7,8 \}$, and $d_3 \in \{1,2\}$. By considering all such $8$  demand vectors, we can list $8$ inequalities in the form of~\eqref{eq:ex2 132 147}, and sum them all together to obtain
\begin{align}
\Rsf  &\geq  \frac{1}{2\Bsf}(|W_{1,\emptyset}|+ |W_{2,\emptyset}| + |W_{4,\emptyset}|+|W_{5,\emptyset}|+|W_{7,\emptyset}| \nonumber\\& +|W_{8,\emptyset}|)  
+ \frac{1}{2\Bsf} (|W_{4,\{2\}}| +|W_{4,\{3\}}| +|W_{5,\{2\}}|+|W_{5,\{3\}}| \nonumber\\& +|W_{7,\{3\}}|  +|W_{8,\{3\}}|    ). \label{eq:ex2 123}
\end{align}
By summing~\eqref{eq:ex2 132} and~\eqref{eq:ex2 123}, we have
\begin{align}
\Rsf &\geq   \sum_{i\in \{1,2,4,5,7,8\}} \left(\frac{1}{2\Bsf} |W_{i,\emptyset}|+  \frac{1}{4\Bsf}   \sum_{j\in \{2,3\}} |W_{i,\{j\}}|  \right) .\label{eq:ex2 perm 1}
\end{align}

 Next we fix one permutation $(u_1,u_2,u_3)=(2,1,3)$, and consider the demand vectors $(d_1,d_2,d_3)$  where     $d_1 \in \{1,2 \}$,   $d_2\in \{4,5 \}$, and $d_3\in \{7,8\}$.
Hence, we can list $8$ inequalities.
 We also fix one permutation $(u_1,u_2,u_3)=(2,3,1)$, and consider the demand vectors $(d_1,d_2,d_3)$  where    $d_1 \in \{4,5\} $,   $d_2\in \{7,8 \}$, and $d_3 \in \{ 1,2 \}$. Hence, we can also list $8$ inequalities.
Then we sum these $16$ inequalities  to obtain
\begin{align}
\Rsf &\geq  \sum_{i\in \{1,2,4,5,7,8\}} \left(\frac{1}{2\Bsf} |W_{i,\emptyset}|+  \frac{1}{4\Bsf}   \sum_{j\in \{1,3\}} |W_{i,\{j\}}|  \right)   .\label{eq:ex2 perm 2}
\end{align}

 Finally we fix one permutation $(u_1,u_2,u_3)=(3,2,1)$, and consider the demand vectors $(d_1,d_2,d_3)$  where     $d_1 \in \{1,2 \}$,   $d_2\in \{4,5 \}$, and $d_3\in \{7,8\}$.
Hence, we can list $8$ inequalities.
 We also fix one permutation $(u_1,u_2,u_3)=(3,1,2)$, and consider the demand vectors $(d_1,d_2,d_3)$  where    $d_1 \in \{4,5\} $,   $d_2\in \{7,8 \}$, and $d_3 \in \{ 1,2 \}$. Hence, we can also list $8$ inequalities.
Then we sum these $16$ inequalities  to obtain
\begin{align}
\Rsf &\geq \sum_{i\in \{1,2,4,5,7,8\}} \left(\frac{1}{2\Bsf} |W_{i,\emptyset}|+  \frac{1}{4\Bsf}   \sum_{j\in \{1,2\}} |W_{i,\{j\}}|  \right)   .\label{eq:ex2 perm 3}
\end{align}

By summing~\eqref{eq:ex2 perm 1}-\eqref{eq:ex2 perm 3}, we obtain
   \begin{subequations}
\begin{align}
\Rsf &\geq \sum_{i\in \{1,2,4,5,7,8\}} \left(\frac{1}{2\Bsf} |W_{i,\emptyset}|+  \frac{1}{6\Bsf}   \sum_{j\in [3]} |W_{i,\{j\}}|  \right) \\& = \frac{1}{2} \alpha_0 +  \frac{1}{6} \alpha_1.\label{eq:ex2 second R}
\end{align}
    \end{subequations}

We take $\frac{2}{3} \times \eqref{eq:recal ex1 final sum ab} + \frac{1}{3} \times \eqref{eq:ex2 second R}$ to obtain
   \begin{subequations}
\begin{align}
\Rsf &\geq \frac{2}{9} \alpha_0 + \frac{2}{9} \beta_0 + \frac{1}{9} \alpha_1  + \frac{1}{6} \alpha_0 +  \frac{1}{18} \alpha_1 \\& = \frac{7}{18}  \alpha_0+ \frac{2}{9} \beta_0+  \frac{1}{6} \alpha_1.  \label{eq:ex2 R}
\end{align}
    \end{subequations}
      Recall that the  file size constraints are given in~\eqref{eq:ex1 file 1} and~\eqref{eq:ex1 file 2}, while the memory size constraint is given in~\eqref{eq:ex1 memory size}.
      From~\eqref{eq:ex1 file 1}, we have
      \begin{align}
  & \frac{7}{18}   \alpha_0 +\frac{7}{18}  \alpha_1+   \frac{7}{18}   \sum_{t_1 \in \{2,3\}} \sum_{i_1 \in \{1,2,4,5,7,8 \} }  \nonumber\\& \sum_{\Tc_1\subseteq [3]:|\Tc_1|=t_1 } \frac{|W_{i_1,\Tc_1}|}{\Bsf} =\frac{7}{3}.\label{eq:ex2 from file 1}
\end{align}
From~\eqref{eq:ex1 file 2}, we have
\begin{align}
\frac{2}{9}\beta_0 +  \frac{2}{9} \sum_{t_2\in [3]} \sum_{i_2\in \{3,6,9 \} } \sum_{\Tc_2\subseteq [3]:|\Tc_2|=t_2} \frac{|W_{i_2,\Tc_2}|}{\Bsf} =\frac{2}{3}.\label{eq:ex2 from file 2}
\end{align}
 From~\eqref{eq:ex1 memory size}, we have
 \begin{align}
& \frac{2}{9}\alpha_1 +  \frac{2}{9}\sum_{t_1 \in \{2,3\}} \sum_{i_1\in \{1,2,4,5,7,8 \} } \sum_{\Tc_1\subseteq [3]:|\Tc_1|=t_1 }    \frac{t_1  |W_{i_1,\Tc_1}|}{\Bsf} \nonumber\\& +
\frac{2}{9} \sum_{t_2\in [3]} \sum_{i_2\in \{3,6,9 \} } \sum_{\Tc_2\subseteq [3]:|\Tc_2|=t_2}  \frac{t_2|W_{i_2,\Tc_2}|}{\Bsf} \leq \frac{2}{3}\Msf.\label{eq:ex2 from memory}
 \end{align}
 By taking $\eqref{eq:ex2 from file 1}+\eqref{eq:ex2 from file 2}-\eqref{eq:ex2 from memory}$, we have
 \begin{align}
   \frac{7}{18}   \alpha_0  + \frac{2}{9}\beta_0  +  \frac{1}{6}\alpha_1 \geq 3-\frac{2}{3}\Msf.\label{eq:ex2 from file memory}
 \end{align}
From~\eqref{eq:ex2 R} and~\eqref{eq:ex2 from file memory}, we have
\begin{align}
\Rsf \geq 3-\frac{2}{3}\Msf. \label{eq:ex final}
\end{align}
Hence, from~\eqref{eq:ex final} we have $\Rsf^{\star}_{\rm u} \geq 3-\frac{2}{3}\Msf$, which    coincides with the achieved load for Theorem~\ref{thm:main result} for $0\leq \Msf \leq 3$.
\hfill $\square$
\end{example}

   We are now ready to generalize the converse proof in Example~\ref{ex:case 1 sec regime} for the case     where $\bsf(\Ksf-1)<2\asf$ and $0  \leq  \Msf \leq \asf+\bsf$.  For any caching scheme with uncoded cache placement ${\bf Z}$,
with the definition of $\alpha_0$, $\beta_0$, and $\alpha_1$ in~\eqref{eq:first regime final sum}, it has been proved in~\eqref{eq:first regime final sum} that
\begin{align}
\Rsf \geq    \frac{\Ksf-1}{\asf \Ksf}  \alpha_0   +  \frac{1}{\bsf \Ksf}  \beta_0
 + \frac{\Ksf-1}{2\asf \Ksf}    \alpha_1  .\label{eq:recal final sum ab}
\end{align}

    Now we     fix one integer $k\in [\Ksf]$. For this integer, we consider two permutations of users, $(k, <k-1>_{\Ksf},\ldots, <k-\Ksf+1>_{\Ksf})$ and
$(k, <k+1>_{\Ksf},\ldots, <k+\Ksf-1>_{\Ksf})$.

For the first permutation $(u_1,u_2,\ldots,u_{\Ksf})=(k, <k-1>_{\Ksf},\ldots, <k-\Ksf+1>_{\Ksf})$, we    consider the demand vectors $(d_1,\ldots,d_{\Ksf})$ where
$d_{u_j}\in\Dc_{u_j,1}$ for $j\in [\Ksf]$, totally $\asf^{\Ksf}$ demand vectors.
 For each $(d_1,\ldots,d_{\Ksf})$, we construct a genie-aided super user with cache as in~\eqref{eq:cache vitrual user} and derive
an inequality as in~\eqref{eq:general form of acyclic first perm}. By considering all such $\asf^{\Ksf}$ demand vectors, we list  $\asf^{\Ksf}$
 inequalities, and sum them all together to obtain
\begin{align}
& \Rsf \geq  \frac{1}{\asf\Bsf} \sum_{i_1\in \Dc_{k,1}} \left(|W_{i_1,\emptyset}|+\sum_{j_1\in [\Ksf]\setminus\{k\}} |W_{i_1, \{j_1\}}|  \right) \nonumber\\& +\frac{1}{\asf\Bsf} \sum_{i_2\in \Dc_{<k-1>_{\Ksf},1}} \left(|W_{i_2,\emptyset}|+\negmedspace\negmedspace\negmedspace\negmedspace \negmedspace\negmedspace\negmedspace\negmedspace \sum_{j_2\in [\Ksf]\setminus\{k, <k-1>_{\Ksf}\}} \negmedspace\negmedspace\negmedspace\negmedspace \negmedspace\negmedspace\negmedspace\negmedspace |W_{i_2, \{j_2\}}|  \right)\nonumber\\ &
+ \cdots +  \frac{1}{\asf \Bsf} \negmedspace \negmedspace \sum_{\substack{i_{\Ksf-1}\in \\  \Dc_{<k-\Ksf+2>_{\Ksf},1}}} \negmedspace \negmedspace \Bigg(|W_{i_{\Ksf-1},\emptyset}|+\negmedspace\negmedspace\negmedspace\negmedspace \negmedspace \sum_{\substack{j_{\Ksf-1}\in [\Ksf]\setminus\{k, \\ <k-1>_{\Ksf},\ldots, <k-\Ksf+2>_{\Ksf} \}}} \negmedspace\negmedspace\negmedspace\negmedspace\negmedspace\negmedspace\negmedspace \negmedspace \nonumber\\& |W_{i_{\Ksf-1}, \{j_{\Ksf-1}\}}|  \Bigg)
  + \frac{1}{\asf \Bsf} \sum_{i_{\Ksf}\in \Dc_{<k-\Ksf+1>_{\Ksf},1}} |W_{i_{\Ksf},\emptyset}|   .\label{eq:general sum form of acyclic first perm sec stra}
\end{align}

For the second permutation $(u_1,u_2,\ldots,u_{\Ksf})=(k, <k+1>_{\Ksf},\ldots, <k+\Ksf-1>_{\Ksf})$, we    consider the demand vectors $(d_1,\ldots,d_{\Ksf})$ where
$d_{u_j}\in\Dc_{u_j,3}=\Dc_{<u_j+1>_{\Ksf},1}$ for $j\in [\Ksf]$, totally $\asf^{\Ksf}$ demand vectors.
For each of such demand vectors, we construct a genie-aided super user with cache as in~\eqref{eq:cache vitrual user} and derive
an inequality as in~\eqref{eq:general form of acyclic first perm}. By summing all the obtained $\asf^{\Ksf}$ inequalities, we have
% \begin{subequations}
\begin{align}
& \Rsf \geq    \frac{1}{\asf\Bsf}    \sum_{i_1\in \Dc_{<k+1>_{\Ksf} ,1}} \negmedspace\negmedspace\negmedspace  \left(|W_{i_1,\emptyset}|+\sum_{j_1\in [\Ksf]\setminus\{k\}} \negmedspace\negmedspace  |W_{i_1, \{j_1\}}|  \right) \nonumber\\& +
\frac{1}{\asf\Bsf}  \negmedspace  \sum_{i_2\in \Dc_{<k+2>_{\Ksf},1}} \negmedspace\negmedspace\negmedspace \left(|W_{i_2,\emptyset}|+ \negmedspace\negmedspace\negmedspace\negmedspace  \sum_{j_2\in [\Ksf]\setminus\{k,<k+1>_{\Ksf}\}}   \negmedspace\negmedspace\negmedspace\negmedspace\negmedspace\negmedspace \negmedspace\negmedspace\negmedspace  |W_{i_2, \{j_2\}}|  \right)
+ \cdots + \nonumber\\ & \frac{1}{\asf \Bsf} \sum_{i_{\Ksf-1}\in \Dc_{<k+\Ksf-1>_{\Ksf},1}} \Bigg(|W_{i_{\Ksf-1},\emptyset}| +\sum_{\substack{j_{\Ksf-1}\in [\Ksf]\setminus\{k,\\ <k+1>_{\Ksf}, \ldots, <k+\Ksf-2>_{\Ksf}\}}}   \negmedspace\negmedspace\nonumber\\& |W_{i_{\Ksf-1}, \{j_{\Ksf-1}\}}|  \Bigg)
  + \frac{1}{\asf \Bsf} \sum_{i_{\Ksf}\in \Dc_{<k+\Ksf>_{\Ksf},1}} |W_{i_{\Ksf},\emptyset}|
  .\label{eq:general sum form of acyclic sec perm sec stra}
\end{align}
 % \end{subequations}

 By summing~\eqref{eq:general sum form of acyclic first perm sec stra} and~\eqref{eq:general sum form of acyclic sec perm sec stra}, we obtain
 \begin{align}
& \Rsf \geq \frac{1}{\asf} \sum_{i_1\in \Cc_1} \frac{|W_{i_1,\emptyset}|}{\Bsf} + \frac{1}{2\asf} \sum_{i_2\in \Cc_1}\sum_{j\in [\Ksf]\setminus \{k\}} \frac{|W_{i_2,\{j\}}|}{\Bsf} .\label{eq:general sum form of acyclic k sec stra}
 \end{align}

 By considering all $k\in [\Ksf]$, we list $\Ksf$ inequalities in the form of~\eqref{eq:general sum form of acyclic k sec stra}, and sum them   to obtain
 \begin{subequations}
 \begin{align}
  \Rsf &\geq \frac{1}{\asf} \sum_{i_1\in \Cc_1} \frac{|W_{i_1,\emptyset}|}{\Bsf} +  \frac{\Ksf-1}{2\asf\Ksf} \sum_{i_2\in \Cc_1}\sum_{j\in [\Ksf]} \frac{|W_{i_2,\{j\}}|}{\Bsf} \\&  =   \frac{1}{\asf} \alpha_0 +  \frac{\Ksf-1}{2\asf\Ksf} \alpha_1.\label{eq:general sum form of acyclic sec stra final}
 \end{align}
 \end{subequations}
Note that,  since  $\bsf(\Ksf-1)<2\asf$, we have
$1-\frac{(\Ksf+1)\bsf}{2(\asf+\bsf)}= \frac{2\asf-\bsf(\Ksf-1)}{2(\asf+\bsf)} > 0$. Hence,
we take $\frac{(\Ksf+1)\bsf}{2(\asf+\bsf)} \times \eqref{eq:recal final sum ab} + \left(1-\frac{(\Ksf+1)\bsf}{2(\asf+\bsf)}\right) \times \eqref{eq:general sum form of acyclic sec stra final} $ to obtain
\begin{align}
\Rsf \geq  \frac{2\asf\Ksf+\bsf(\Ksf-1)}{2(\asf+\bsf)\asf\Ksf} \alpha_0 + \frac{\Ksf+1}{2(\asf+\bsf)\Ksf} \beta_0 +\frac{\Ksf-1}{2\asf \Ksf}    \alpha_1 .\label{eq:sec regime lower bound on R}
\end{align}
Recall that the  file size constraints are given in~\eqref{eq:first regime file 1} and~\eqref{eq:first regime file 2}, while the memory size constraint is given in~\eqref{eq:first regime memory size}. By taking $\frac{2\asf\Ksf+\bsf(\Ksf-1)}{2(\asf+\bsf)\asf\Ksf} \times \eqref{eq:first regime file 1} + \frac{\Ksf+1}{2(\asf+\bsf)\Ksf} \times \eqref{eq:first regime file 2} - \frac{\Ksf+1}{2(\asf+\bsf)\Ksf} \times \eqref{eq:first regime memory size}$, we obtain
 \begin{subequations}
\begin{align}
&\frac{2\asf\Ksf+\bsf(\Ksf-1)}{2(\asf+\bsf)\asf\Ksf} \alpha_0 + \frac{\Ksf+1}{2(\asf+\bsf)\Ksf} \beta_0 +
%\left(\frac{2\asf\Ksf+\bsf(\Ksf-1)}{2(\asf+\bsf)\asf\Ksf}-  \frac{\Ksf+1}{2(\asf+\bsf)\Ksf} \right)
   \frac{\Ksf-1}{2\asf\Ksf}\alpha_1 \nonumber\\
   & \geq \Ksf -\frac{\Ksf+1}{2(\asf+\bsf)}\Msf + \frac{1}{2(\asf+\bsf)\asf\Ksf} \sum_{t_1 \in [2:\Ksf]} \nonumber\\
   & \sum_{i_1\in \Cc_1 } \sum_{\Tc_1\subseteq [\Ksf]:|\Tc_1|=t_1 }  \left( (t_1-2)\asf\Ksf+t_1 \asf -\bsf(\Ksf-1)\right) \frac{|W_{i_1,\Tc_1}|}{\Bsf} \nonumber\\
   &+   \frac{\Ksf+1}{2(\asf+\bsf)\Ksf} \sum_{t_2\in [\Ksf]} \sum_{i_2\in \Cc_2 } \sum_{\Tc_2\subseteq [\Ksf]:|\Tc_2|=t_2} (t_2-1)\frac{|W_{i_2,\Tc_2}|}{\Bsf} \\
  & \geq \Ksf -\frac{\Ksf+1}{2(\asf+\bsf)}\Msf,\label{eq:sec regime final}
\end{align}
  \end{subequations}
  where~\eqref{eq:sec regime final} comes from that $\bsf(\Ksf-1)<2\asf$. From~\eqref{eq:sec regime lower bound on R} and~\eqref{eq:sec regime final},  we have
\begin{align}
\Rsf  \geq \Ksf -\frac{\Ksf+1}{2(\asf+\bsf)}\Msf,
\end{align}
  which leads to $\Rsf^{\star}_{\rm u} \geq \Ksf -\frac{\Ksf+1}{2(\asf+\bsf)}\Msf$, coinciding with the achieved load for Theorem~\ref{thm:main result} when $\bsf(\Ksf-1)<2\asf$  and $0\leq  \Msf \leq  \asf+\bsf$.
\subsection{Converse Proof of Theorem~\ref{thm:main result}: $\bsf(\Ksf-1)\geq 2\asf$}
\label{sec:converse thrid seg}
In the end,  we   focus on the case where $\bsf(\Ksf-1) \geq 2\asf$.
For any caching scheme with uncoded cache placement ${\bf Z}$,  we will first prove that
\begin{align}
\Rsf \geq  \frac{1}{\bsf} \sum_{i\in \Cc_2} \frac{|W_{i,\emptyset}|}{\Bsf} = \frac{1}{\bsf} \beta_0, \label{eq:R lower bound on beta}
\end{align}
where $\Cc_2 $ and $\beta_0$ are    defined in~\eqref{eq:def of C2} and~\eqref{eq:first regime final sum}, respectively.

More precisely, each time we consider a demand vector $(d_1,\ldots,d_{\Ksf})$, where $d_k \in \Dc_{k,2}$ for each $k\in [\Ksf]$. Since $ W_{d_1,\emptyset},\ldots,W_{d_{\Ksf},\emptyset}$ are   not cached by any cache node and from $(Z_1,\ldots, Z_{\Ksf}, X)$ we can recover $W_{d_1,\emptyset},\ldots,W_{d_{\Ksf},\emptyset}$,    we have
  \begin{subequations}
\begin{align}
&\Rsf \geq \frac{H(X) }{\Bsf} \geq \frac{H(X | Z_1,\ldots, Z_{\Ksf} )}{\Bsf} \\
& =   \frac{ H(X,  W_{d_1,\emptyset},\ldots,W_{d_{\Ksf},\emptyset}| Z_1,\ldots, Z_{\Ksf} )}{\Bsf} \\
& \geq  \frac{H(   W_{d_1,\emptyset},\ldots,W_{d_{\Ksf},\emptyset}| Z_1,\ldots, Z_{\Ksf} )}{\Bsf}  \\& = \frac{H(   W_{d_1,\emptyset},\ldots,W_{d_{\Ksf},\emptyset}) }{\Bsf} \\
& = \frac{| W_{d_1,\emptyset}|}{\Bsf}   + \cdots +  \frac{ | W_{d_{\Ksf},\emptyset}|}{\Bsf} . \label{eq:R larger than emp}
\end{align}
  \end{subequations}
  By considering all demand vectors $(d_1,\ldots,d_{\Ksf})$, where $d_k \in \Dc_{k,2}$ for each $k\in [\Ksf]$ , we list $\bsf^{\Ksf}$ inequalities in the form of~\eqref{eq:R larger than emp}, and sum them all together to obtain~\eqref{eq:R lower bound on beta}.

  In the following, we will   prove the converse bound for case where $\bsf(\Ksf-1) \geq 2\asf$, by the help of the derived lower bounds of $\Rsf$ in~\eqref{eq:first regime final sum} and~\eqref{eq:R lower bound on beta}, the file constraints  in~\eqref{eq:first regime file 1} and~\eqref{eq:first regime file 2}, and the memory size constraint  in~\eqref{eq:first regime memory size}.

More precisely, since $\bsf(\Ksf-1) \geq 2\asf$, we have $1-\frac{2\asf\Ksf}{(\Ksf-1)(2\asf+\bsf)}=\frac{\bsf(\Ksf-1)-2\asf}{(\Ksf-1)(2\asf+\bsf)}\geq 0$. Hence, by taking $\frac{2\asf\Ksf}{(\Ksf-1)(2\asf+\bsf)} \times \eqref{eq:first regime final sum} + \left( 1-\frac{2\asf\Ksf}{(\Ksf-1)(2\asf+\bsf)} \right) \times \eqref{eq:R lower bound on beta}$, we obtain
\begin{align}
\Rsf \geq  \frac{2}{2\asf+\bsf} \alpha_0 + \frac{1}{2\asf+\bsf} \beta_0+ \frac{1}{2\asf+\bsf }\alpha_1.\label{eq:third regime lower bound on R}
%\frac{\Ksf-1}{\asf \Ksf}  \frac{2\asf\Ksf}{(\Ksf-1)(2\asf+\bsf)}  \alpha_0 + ( \frac{1}{\bsf \Ksf} \frac{2\asf\Ksf}{(\Ksf-1)(2\asf+\bsf)} + \frac{1}{\bsf}\frac{\bsf(\Ksf-1)-2\asf}{(\Ksf-1)(2\asf+\bsf)}) \beta_0 +  \frac{\Ksf-1}{2\asf \Ksf}   \frac{2\asf\Ksf}{(\Ksf-1)(2\asf+\bsf)}    \alpha_1  .
\end{align}
 Next, by taking $\frac{2}{2\asf+\bsf} \times \eqref{eq:first regime file 1} + \frac{1}{2\asf+\bsf} \times \eqref{eq:first regime file 2}- \frac{1}{2\asf+\bsf} \times \eqref{eq:first regime memory size}$, we obtain
   \begin{subequations}
 \begin{align}
&  \frac{2}{2\asf+\bsf} \alpha_0 +\frac{1}{2\asf+\bsf} \beta_0 + \frac{1}{2\asf+\bsf} \alpha_1 \nonumber\\
& \geq \Ksf-\frac{\Ksf}{2\asf+\bsf}\Msf +  \frac{1}{2\asf+\bsf} \sum_{t_1 \in [2:\Ksf]} \sum_{i_1\in \Cc_1 } \sum_{\Tc_1\subseteq [\Ksf]:|\Tc_1|=t_1 } (t_1-2) \nonumber\\
& \frac{|W_{i_1,\Tc_1}|}{\Bsf}  +  \frac{1}{2\asf+\bsf} \sum_{t_2\in [\Ksf]} \sum_{i_2\in \Cc_2 } \sum_{\Tc_2\subseteq [\Ksf]:|\Tc_2|=t_2} (t_2-1)\frac{|W_{i_2,\Tc_2}|}{\Bsf} \\
& \geq  \Ksf-\frac{\Ksf}{2\asf+\bsf}\Msf.\label{eq:third regime from file memory}
 \end{align}
   \end{subequations}
 From~\eqref{eq:third regime lower bound on R} and~\eqref{eq:third regime from file memory}, we have
 \begin{align}
 \Rsf \geq \Ksf-\frac{\Ksf}{2\asf+\bsf}\Msf,
 \end{align}
 which leads to  $\Rsf^{\star}_{\rm u} \geq \Ksf-\frac{\Ksf}{2\asf+\bsf}\Msf$, coinciding with the achieved load for Theorem~\ref{thm:main result} when $\bsf(\Ksf-1)\geq 2\asf$.

 \section{Extension to Multiaccess Coded Caching Systems}
\label{sec:extension}
\begin{figure}
    \centering
 \includegraphics[scale=0.19]{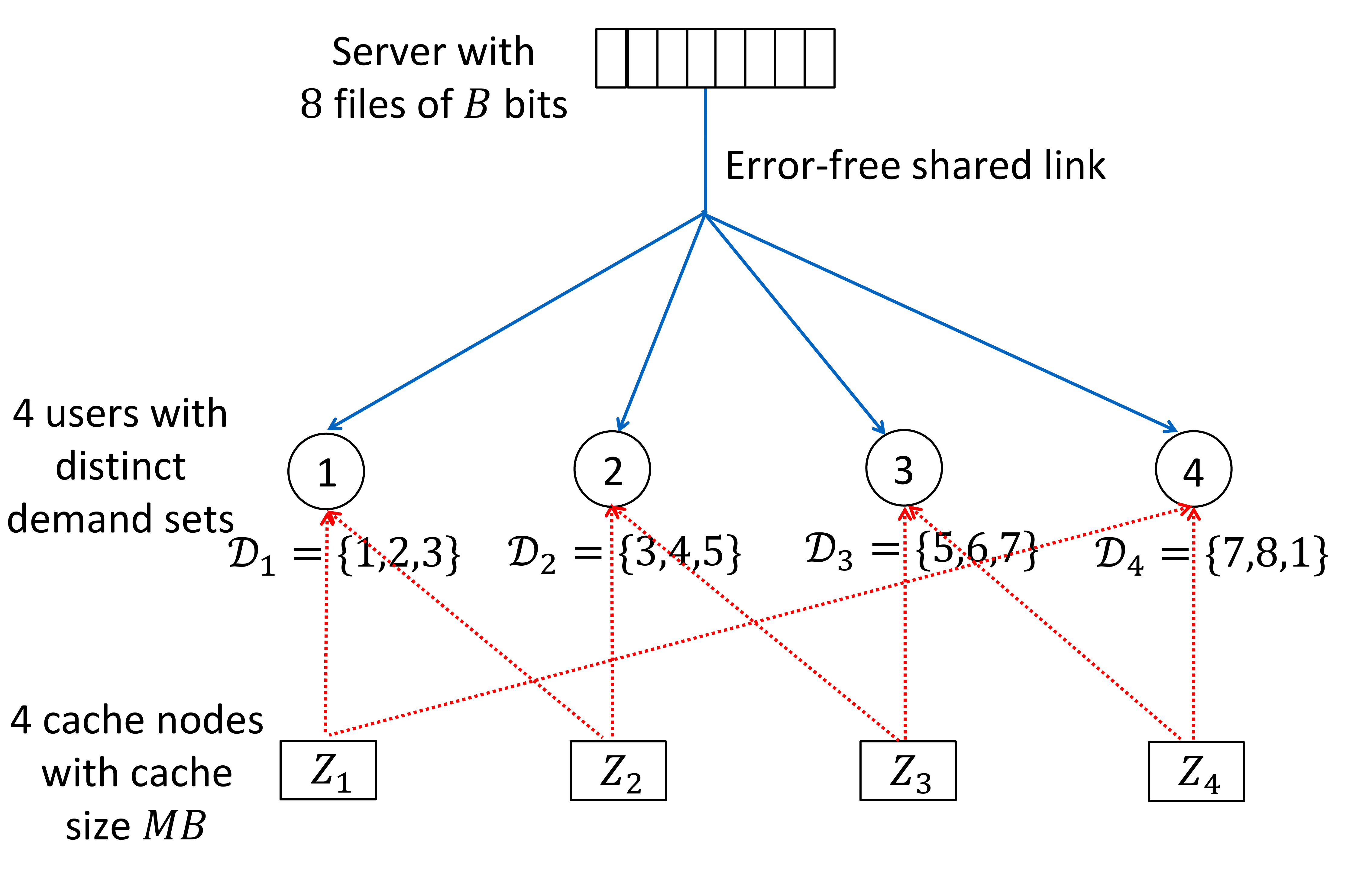}
\caption{\small  The information theoretic model of the  multiaccess   coded caching  problem  for location-based content   with $\Ksf=4$, $\Nsf=8$,   $\asf=\bsf=1$, $\Lsf=2$.}
\label{fig:VNET multiaccess math}
%\vspace{-5mm}
\end{figure}

 An ideal coded edge caching model with a line multiaccess topology, referred to as multiaccess  coded caching, was originally introduced in~\cite{Hachem2017multiaccess}.
Different from our considered edge caching topology in Section~\ref{sec:model} where each user is connected to the nearest cache node,   in~\cite{Hachem2017multiaccess}  each user is connected to  $\Lsf \in [\Ksf]$ cache nodes in a cyclic wrap-around fashion. Since~\cite{Hachem2017multiaccess}, the multiaccess coded caching problem was widely considered, where different achievable and converse bounds were proposed in~\cite{Reddy2020Multiaccess,sasi2020multiaccess,Serbetci2019multiaccess,Cheng2020multiaccess,Reddy2020structured,Ozfatura2020mobility}, while the optimality remains open (even for the uncoded cache placement).
In the following, we consider the $(\Ksf,\asf,\bsf,\Lsf)$ multiaccess coded caching problem for location-based content, as illustrated in  Fig.~\ref{fig:VNET multiaccess math}.  The only difference from the considered system model in      Section~\ref{sec:model} is that,
each user $k\in [\Ksf]$ is connected to the $\Lsf$ cache nodes in $\{k, <k+1>_{\Ksf},\ldots, <k+\Lsf-1>_{\Ksf}\}$, and can retrieve the cached content  of its connected cache nodes without any cost. When $\Lsf=1$, it reduces to the system model in    Section~\ref{sec:model}. Hence, in this section, we consider $\Lsf \in [2:\Ksf]$.
Note that, different from the multiaccess  coded caching problem in~\cite{Hachem2017multiaccess} where each user may request any file in the library, in the considered problem  the set of possible demanded files by  user $k$  is $\Dc_{k}$, where $\Dc_{k}$ is defined in~\eqref{eq:Dk}.

Our objective in the $(\Ksf,\asf,\bsf,\Lsf)$ multiaccess coded caching problem for location-based content   is to design the cache placement and delivery phases, such that the  worst-case load  among all possible demands is minimized, where the optimal worst-case load is denoted by $\Rsf^{\star}$.

We characterize the exact optimality for the considered problem in the following theorem.
\begin{thm}
\label{thm:extension}
For the $(\Ksf,\asf,\bsf,\Lsf)$ multiaccess coded caching problem for location-based content   where $\Lsf\in [2:\Ksf]$, we have
\begin{align}
\Rsf^{\star}= \begin{cases}  \Ksf-\frac{\Ksf }{ \asf+\bsf }\Msf , & \text{ if } \  0 \leq \Msf \leq \asf +\bsf ; \\  0, & \text{ if } \  \Msf \geq \asf+\bsf. \end{cases}
\label{eq:extension}
\end{align}
\hfill $\square$
\end{thm}
 \begin{IEEEproof}

 {\it Achievability.}
When  $\Msf=0$, obviously we have $\Rsf=\Ksf$. In the following we will show that the memory-load tradeoff $(\Msf, \Rsf)=(\asf+\bsf,0)$ is achievable. By the memory sharing between $(0,\Ksf)$ and $(\asf+\bsf,0)$, we can achieve $\Rsf=\Ksf-\frac{\Ksf }{ \asf+\bsf }\Msf$ when $0 \leq \Msf \leq \asf +\bsf$, which coincides with~\eqref{eq:extension}.

 Let us focus on $\Msf=\asf+\bsf$.
In the cache placement phase,  each cache node $k \in [\Ksf]$ caches $W_{n}$ where  $n\in \Dc_{k,1}\cup \Dc_{k,2}$. Since  $ | \Dc_{k,1}\cup \Dc_{k,2}|=\asf+\bsf $, the memory size constraint is satisfied.

In the delivery phase, user $k\in [\Ksf]$ requests $W_{d_k}$ where $d_k\in \Dc_k$. By~\eqref{eq:Dk}, we have $\Dc_k= \Dc_{k,1}\cup\Dc_{k,2}\cup\Dc_{k,3}$ and $\Dc_{k,3}=\Dc_{<k+1>_{\Ksf},1}$. Hence, $\Dc_k= \Dc_{k,1}\cup\Dc_{k,2}\cup \Dc_{<k+1>_{\Ksf},1}$.
If $\Dc_k \in \Dc_{k,1}\cup\Dc_{k,2}$, user $k$ can retrieve $W_{d_k}$ from cache node $k$; otherwise, user $k$ can retrieve $W_{d_k}$ from cache node $<k+1>_{\Ksf}$.
Hence, when $\Msf=\asf+\bsf$, we achieve $\Rsf=0$.

  {\it Converse.}
  Let us focus on the regime $0\leq \Msf \leq \asf+\bsf$.
  For any achievable scheme with the memory-load tradeoff $(\Msf,\Rsf)$,
we consider a cut of all $\Ksf$ cache nodes  and   $\Ksf$   users.
Recall that $\Dc_{k}(i)$ denotes the  $i^{\text{th}}$ smallest element in $\Dc_{k}$.
  For each $i\in [\asf+\bsf]$,  we assume that $X_i$ is transmitted by the  server for the demand vector $(\Dc_{1}(i), \Dc_{2}(i),\ldots, \Dc_{\Ksf  }(i) )$.
Note that
\begin{subequations}
\begin{align}
&\cup_{i\in [\asf+\bsf]} \cup_{k\in [\Ksf]}  \{\Dc_{k}(i)\}  = \left( \cup_{i_1\in [\asf]} \cup_{k_1\in [\Ksf]}  \{\Dc_{k_1}(i_1)\}\right) \nonumber\\& \cup \left( \cup_{i_2\in [\asf+1:\asf+\bsf]} \cup_{k_2\in [\Ksf]}  \{\Dc_{k_2}(i_2)\}\right) \\
& = \left( \cup_{k_1\in [\Ksf]} \Dc_{k_1,1} \right) \cup \left( \cup_{k_1\in [\Ksf]} \Dc_{k_1,2} \right) \\& =\Cc_1 \cup \Cc_2 =[\Nsf],
\end{align}
\end{subequations}
where $\Cc_1$ and $\Cc_2$ are defined in~\eqref{eq:def of C1} and~\eqref{eq:def of C2}, respectively.
Hence, from $(Z_1,\ldots,Z_{\Ksf},X_1,\ldots, X_{\asf+\bsf})$, we can decode $W_{n}$ for each $n\in [\Nsf]$; thus  (recall that $\Nsf:=(\asf+\bsf)\Ksf$)
\begin{subequations}
\begin{align}
&  \Ksf \Msf \Bsf+ ( \asf+\bsf) \Rsf\Bsf \geq   H(Z_1,\ldots,Z_{\Ksf},X_1,\ldots, X_{\asf+\bsf})  \\& \geq    H(W_1,\ldots,W_{\Nsf}),\\
 & \Longrightarrow  \Rsf \geq   \Ksf - \frac{\Ksf}{ \asf+\bsf }\Msf, \\
 & \Longrightarrow  \Rsf^{\star} \geq    \Ksf - \frac{\Ksf}{ \asf+\bsf }\Msf, \label{eq:extension converse}
\end{align}
\end{subequations}
which coincides with~\eqref{eq:extension}.
  \end{IEEEproof}

\begin{rem}
\label{rem:independent of L}
From Theorem~\ref{thm:extension}, it can be seen that when $\Lsf \in [2:\Ksf]$, the optimal load for the considered multiaccess coded caching problem for location-based content   does not depend on $\Lsf$. In other words, 
allowing the users to access more than just their two nearest   local caches does not reduce the load of the common bottleneck link.
%it does not reduce the load if each user is allowed to access more caches.
\hfill $\square$
\end{rem}

\section{Conclusions}
\label{sec:conclusion}
This paper introduced a novel coded caching problem for location-based content    %in the vehicular networks with edge cache nodes. 
  in networks equipped with edge caching nodes. This is motivated, for example, by a vehicular network where self-driving vehicles need to access super High-Definition maps of the region through which they are driving. 
In the proposed model, each user is connected to the nearest cache node and  requests a file in a subset of library depending on its location.
Novel information theoretic converse bounds (with or without the constraint of uncoded cache placement) and achievable scheme were proposed, from which we can show the exact optimality on the worst-case load under uncoded cache placement and the general order optimality within a factor of $3$. We also extended the   coded caching problem for location-based content   to the multiaccess coded caching topology, and characterized the exact optimality if each user is connected to at least two nearest cache nodes.
On-going works  include  characterizing the exact optimality without the constraint of uncoded cache placement, considering the model where each region contains more than one users, and 
studying   two-dimensional  vehicular networks such as     the Manhattan topology.

\appendices

\section{Proof of Theorem~\ref{thm:order optimal}}
\label{sec:converse order optimal}
We divide the proof into two cases, $\Ksf$ is even and $\Ksf$ is odd, respectively.   For each case, we  first propose a general cut-set bound on the optimal load for the considered problem, and then upper bound the multiplicative gap between $\Rsf^{\star}_{\rm u} $ and this cut-set converse.

\subsection{$\Ksf$ is Even}
\label{sub:K is even}
For any achievable scheme with the memory-load tradeoff $(\Msf,\Rsf)$,
we consider a cut of $\frac{\Ksf}{2}$ cache nodes with the indices in  $\Vc= \{1,3, \ldots, \Ksf -1 \}$, and their connected users.   It can be seen that  for any $k_1 \neq k_2$ and $k_1,k_2 \in \Vc$, we have $\Dc_{k_1} \cap \Dc_{k_2}=\emptyset$. Denote the $i^{\text{th}}$ smallest element in $\Dc_{k}$ by $\Dc_{k}(i)$, where $i \in [2\asf+\bsf]$ and $k\in [\Ksf/2]$.
For each $i\in [2\asf+\bsf]$, we assume that $X_i$ is transmitted by the  server for the demand vector $(\Dc_{1}(i), \Dc_{3}(i),\ldots, \Dc_{\Ksf -1}(i) )$.
From $(Z_1, Z_3, \ldots, Z_{\Ksf-1}, X_1,X_2,\ldots, X_{2\asf+\bsf})$, we can decode $W_{n}$ where $n\in \cup_{k\in\Vc } \Dc_{k}$; thus by the cut-set bound,
%\begin{subequations}
\begin{align}
& \frac{\Ksf}{2}\Msf \Bsf+ (2\asf+\bsf) \Rsf\Bsf \nonumber\\& \geq   H(Z_1, Z_3, \ldots, Z_{\Ksf-1}, X_1,X_2,\ldots, X_{2\asf+\bsf})   \nonumber\\& \geq    H\big( (W_n:n\in \cup_{k\in\Vc } \Dc_{k})  \big), \nonumber \\
 & \Longrightarrow  \Rsf \geq   \frac{\Ksf}{2}- \frac{\Ksf}{2(2\asf+\bsf)}\Msf, \nonumber \\
 & \Longrightarrow  \Rsf^{\star} \geq   \frac{\Ksf}{2}- \frac{\Ksf}{2(2\asf+\bsf)}\Msf. \label{eq:first case cut set}
\end{align}
%\end{subequations}

Let us then compare the  converse bound in~\eqref{eq:first case cut set} with the   achievable bound in Theorem~\ref{thm:main result}.

First, we consider   $\bsf(\Ksf-1)<2\asf$ and $0 \leq \Msf \leq \asf+\bsf$, for which the achieved load is $\Rsf^{\star}_{\rm u}=\Ksf-\frac{\Ksf+1}{2(\asf+\bsf)}\Msf$.
In this regime, the converse bound in~\eqref{eq:first case cut set} is the memory sharing between $(0, \Ksf/2)$ and  $\left(\asf+\bsf, \frac{\Ksf}{2}-\frac{\Ksf(\asf+\bsf)}{2(2\asf+\bsf)} \right)$, while the achievable bound is the    memory sharing between $(0, \Ksf )$ and  $\left(\asf+\bsf, \frac{\Ksf-1}{2}  \right)$.
When $\Msf=0$, the multiplicative gap between the achievable bound and the converse bound is $2$. When $\Msf=\asf+\bsf$,    the multiplicative gap between the achievable bound and the converse bound is within a factor of $2$; this is because
\begin{subequations}
\begin{align}
2 \left( \frac{\Ksf}{2}-\frac{\Ksf(\asf+\bsf)}{2(2\asf+\bsf)} \right) -  \frac{\Ksf-1}{2} &  = \frac{\asf \Ksf}{2\asf +\bsf}-\frac{\Ksf-1}{2}  \\&  = \frac{2\asf -\bsf(\Ksf-1) }{2(2\asf+\bsf)} >0.
\end{align}
\end{subequations}
Hence, when $\bsf(\Ksf-1)<2\asf$ and $0 \leq \Msf \leq \asf+\bsf $,  we can prove $2 \Rsf^{\star} \geq \Rsf^{\star}_{\rm u}$.

Second, we consider $\bsf(\Ksf-1)<2\asf$ and $ \asf+\bsf \leq  \Msf \leq 2\asf+\bsf$,  for which the achieved load is $\Rsf^{\star}_{\rm u}=\frac{(\Ksf-1)(2\asf+\bsf)}{2\asf}-\frac{\Ksf-1}{2\asf}\Msf$.  In this regime, the converse bound in~\eqref{eq:first case cut set} is the memory sharing between   $\left(\asf+\bsf, \frac{\Ksf}{2}-\frac{\Ksf(\asf+\bsf)}{2(2\asf+\bsf)} \right)$ and $(2\asf+\bsf, 0 )$, while the achievable bound is the    memory sharing between $\left(\asf+\bsf, \frac{\Ksf-1}{2}  \right)$ and  $(2\asf+\bsf, 0 )$. It has been proved that  when $\Msf=\asf+\bsf$,    the multiplicative gap between the achievable bound and the converse bound is within a factor of $2$. In addition, when $\Msf=2\asf+\bsf$, the achievable bound coincides with the converse bound. Hence, when $\bsf(\Ksf-1)<2\asf$ and $ \asf+\bsf \leq  \Msf \leq 2\asf+\bsf$,  we can prove $2 \Rsf^{\star} \geq \Rsf^{\star}_{\rm u}$.

Third, we consider $\bsf(\Ksf-1)\geq 2\asf$,  for which the achieved load is $\Rsf^{\star}_{\rm u}= \Ksf-\frac{\Ksf}{2\asf+\bsf}\Msf$.
When $ 0 \leq  \Msf \leq 2\asf+\bsf$,  the converse bound in~\eqref{eq:first case cut set} is the memory sharing between   $\left(0, \Ksf/2 \right)$ and $(2\asf+\bsf, 0 )$, while the achievable bound is the    memory sharing between $\left(0,  \Ksf \right)$ and  $(2\asf+\bsf, 0 )$.
Hence,  in this case, we have $2 \Rsf^{\star} \geq \Rsf^{\star}_{\rm u}$.

In conclusion, when $\Ksf$ is even, the proposed scheme for Theorem~\ref{thm:main result} is generally order optimal within a constant of $2$.

\iffalse
 From~\eqref{eq:first case cut set}, we have
\begin{subequations}
\begin{align}
& 3 \Rsf^{\star} - \Rsf^{\star}_{\rm u} \geq   \frac{3\Ksf}{2}- \frac{3\Ksf}{2(2\asf+\bsf)}\Msf -\Ksf+\frac{\Ksf+1}{2(\asf+\bsf)}\Msf \\
& = \frac{\Ksf}{2} - \frac{(\asf+2\bsf)\Ksf}{2(\asf+\bsf)(2\asf+\bsf)} \Msf \\
&\geq \frac{\Ksf}{2} -  \frac{(\asf+2\bsf)\Ksf}{2 (2\asf+\bsf)}  \\
\end{align}
\end{subequations}
\fi
\subsection{$\Ksf$ is Odd}
\label{sub:K is odd}
In the following, we  consider  $\Ksf$ is odd and  $\Ksf \geq 3$.

For any achievable scheme with the memory-load tradeoff $(\Msf,\Rsf)$,
we consider a cut of $\frac{\Ksf-1}{2}$ cache nodes with the indices in  $\Vc= \{1,3, \ldots, \Ksf -2 \}$, and their connected users.   It can be seen that  for any $k_1 \neq k_2$ and $k_1,k_2 \in \Vc$, we have $\Dc_{k_1} \cap \Dc_{k_2}=\emptyset$.
For each $i\in [2\asf+\bsf]$, we assume that $X_i$ is transmitted by the  server for the demand vector $(\Dc_{1}(i), \Dc_{3}(i),\ldots, \Dc_{\Ksf -2 }(i) )$.
From $(Z_1, Z_3, \ldots, Z_{\Ksf -2 }, X_1,X_2,\ldots, X_{2\asf+\bsf})$, we can decode $W_{n}$ where $n\in \cup_{k\in\Vc } \Dc_{k}$; thus by the cut-set bound,
%\begin{subequations}
\begin{align}
& \frac{\Ksf-1}{2} \Msf \Bsf+ (2\asf+\bsf) \Rsf\Bsf \nonumber\\ &\geq   H(Z_1, Z_3, \ldots, Z_{\Ksf-2}, X_1,X_2,\ldots, X_{2\asf+\bsf})  \nonumber\\& \geq    H\big( (W_n:n\in \cup_{k\in\Vc } \Dc_{k})  \big),  \nonumber\\
 & \Longrightarrow  \Rsf \geq   \frac{\Ksf-1}{2}- \frac{\Ksf-1}{2(2\asf+\bsf)}\Msf,  \nonumber\\
 & \Longrightarrow  \Rsf^{\star} \geq    \frac{\Ksf-1}{2}- \frac{\Ksf-1}{2(2\asf+\bsf)}\Msf. \label{eq:sec case cut set}
\end{align}
%\end{subequations}

We also  compare the  converse bound in~\eqref{eq:sec case cut set} with the   achievable bound in Theorem~\ref{thm:main result}.

First, we consider   $\bsf(\Ksf-1)<2\asf$ and $0 \leq \Msf \leq \asf+\bsf$, for which the achieved load is $\Rsf^{\star}_{\rm u}=\Ksf-\frac{\Ksf+1}{2(\asf+\bsf)}\Msf$.
In this regime, the converse bound in~\eqref{eq:sec case cut set} is the memory sharing between $\left(0, \frac{\Ksf-1}{2} \right)$ and  $\left(\asf+\bsf, \frac{\Ksf-1}{2}-\frac{(\Ksf-1)(\asf+\bsf)}{2(2\asf+\bsf)} \right)$, while the achievable bound is the    memory sharing between $(0, \Ksf )$ and  $\left(\asf+\bsf, \frac{\Ksf-1}{2}  \right)$.
When $\Msf=0$, the multiplicative gap between the achievable bound and the converse bound is within a factor $\frac{2\Ksf}{\Ksf-1} \leq 3$. When $\Msf=\asf+\bsf$,    the multiplicative gap between the achievable bound and the converse bound is within a factor of $3$; this is because
%\begin{subequations}
\begin{align}
& 3 \left( \frac{\Ksf-1}{2}-\frac{(\Ksf-1)(\asf+\bsf)}{2(2\asf+\bsf)} \right) -  \frac{\Ksf-1}{2} \nonumber\\& = (\Ksf-1) \left(  1-   \frac{ 3(\asf+\bsf)}{ 2(2\asf+\bsf)} \right)  =  (\Ksf-1) \frac{\asf-\bsf}{2(2\asf+\bsf) } \nonumber \\
& >0, \label{eq:K odd first regime}
\end{align}
%\end{subequations}
where~\eqref{eq:K odd first regime} comes from    $\bsf(\Ksf-1)<2\asf$ and $\Ksf \geq 3$, which leads  to $\asf >\bsf$.
Hence, when $\bsf(\Ksf-1)<2\asf$ and $0 \leq \Msf \leq \asf+\bsf$,  we can prove $3 \Rsf^{\star} \geq \Rsf^{\star}_{\rm u}$.

Second, we consider $\bsf(\Ksf-1)<2\asf$ and $ \asf+\bsf \leq  \Msf \leq 2\asf+\bsf$,  for which the achieved load is $\Rsf^{\star}_{\rm u}=\frac{(\Ksf-1)(2\asf+\bsf)}{2\asf}-\frac{\Ksf-1}{2\asf}\Msf$.  In this regime, the converse bound in~\eqref{eq:sec case cut set} is the memory sharing between   $\left(\asf+\bsf, \frac{\Ksf-1}{2}-\frac{(\Ksf-1)(\asf+\bsf)}{2(2\asf+\bsf)} \right)$ and $(2\asf+\bsf, 0 )$, while the achievable bound is the    memory sharing between $\left(\asf+\bsf, \frac{\Ksf-1}{2}  \right)$ and  $(2\asf+\bsf, 0 )$. It has been proved that  when $\Msf=\asf+\bsf$,    the multiplicative gap between the achievable bound and the converse bound is within a factor of $3$. In addition, when $\Msf=2\asf+\bsf$, the achievable bound coincides with the converse bound. Hence, when $\bsf(\Ksf-1)<2\asf$ and $ \asf+\bsf \leq  \Msf \leq 2\asf+\bsf$,  we can prove $3\Rsf^{\star} \geq \Rsf^{\star}_{\rm u}$.

Third, we consider $\bsf(\Ksf-1)\geq 2\asf$,  for which the achieved load is $\Rsf^{\star}_{\rm u}= \Ksf-\frac{\Ksf}{2\asf+\bsf}\Msf$.
When $ 0 \leq  \Msf \leq 2\asf+\bsf$,  the converse bound in~\eqref{eq:sec case cut set} is the memory sharing between   $\left(0, \frac{\Ksf-1}{2} \right)$ and $(2\asf+\bsf, 0 )$, while the achievable bound is the    memory sharing between $\left(0,  \Ksf \right)$ and  $(2\asf+\bsf, 0 )$.
It has been proved that  when $\Msf=0$,    the multiplicative gap between the achievable bound and the converse bound is within a factor of $3$. In addition, when $\Msf=2\asf+\bsf$, the achievable bound coincides with the converse bound.
Hence,  in this case, we have $3 \Rsf^{\star} \geq \Rsf^{\star}_{\rm u}$.

In conclusion, when $\Ksf$ is even, the proposed scheme for Theorem~\ref{thm:main result} is generally order optimal within a constant of $3$.

\bibliographystyle{IEEEtran}
\bibliography{IEEEabrv,IEEEexample}

\begin{IEEEbiography}
 [{\includegraphics[width=1in,height=1.25in,clip,keepaspectratio]{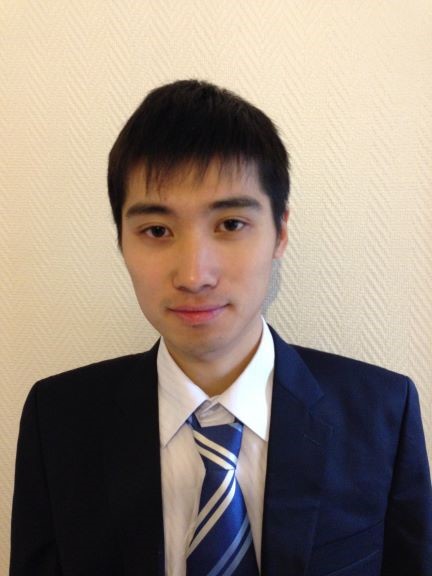}}]{Kai Wan} (S'15 -- M'18) received  the B.E. degree in    Optoelectronics from  Huazhong University of Science and Technology, China, in 2012, the   M.Sc. and Ph.D. degrees in Communications from Universit{\'e}  Paris-Saclay, France, in 2014 and 2018.  He is currently a post-doctoral    researcher with the Communications and Information Theory Chair   (CommIT) at Technische Universit\"at Berlin, Berlin, Germany. His   research interests include information theory, coding techniques, and   their applications on coded caching,  index coding, distributed storage,  distributed computing, wireless communications,   privacy and security. He has served as an Associate Editor of IEEE Communications Letters from Aug. 2021.
\end{IEEEbiography}

 \begin{IEEEbiography}
[{\includegraphics[width=1in,height=1.25in,clip,keepaspectratio]{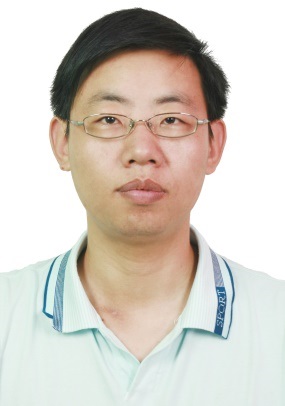}}]{Minquan Cheng}
 received the Ph.D. degree from Department of Social Systems and Management, Graduate School of Systems and Information Engineering, University of Tsukuba, Tsukuba, Ibaraki, Japan, in 2012. Then, he joined Guangxi Normal University, Guilin, Guangxi, China, where he is currently a full professor at School of Computer Science and Information Technology. His research interests include combinatorics, coding theory, cryptography
 and their interactions.
\end{IEEEbiography}

\begin{IEEEbiography}
 [{\includegraphics[width=1in,height=1.25in,clip,keepaspectratio]{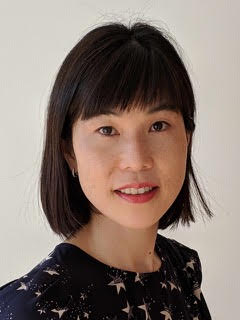}}]{Mari Kobayashi}  (M'06–SM'15) received the B.E. degree in electrical engineering from Keio University, Yokohama, Japan, in 1999, and the M.S. degree in mobile radio and the Ph.D. degree from École Nationale Supérieure des Télécommunications, Paris, France, in 2000 and 2005, respectively. From November 2005 to March 2007, she was a postdoctoral researcher at the Centre Tecnològic de Telecomunicacions de Catalunya, Barcelona, Spain. In May 2007, she joined the Telecommunications department at Centrale Supélec, Gif-sur-Yvette, France, where she is now a professor. She is the recipient of the Newcom++ Best Paper Award in 2010, and IEEE Comsoc/IT Joint Society Paper Award in 2011, and ICC Best Paper Award in 2019. She was an Alexander von Humboldt Experienced Research Fellow (September 2017- April 2019) and an August-Wihelm Scheer Visiting Professor (August 2019-April 2020) at Technical University of Munich (TUM).  

\end{IEEEbiography}

\begin{IEEEbiography}
[{\includegraphics[width=1in,height=1.25in,clip,keepaspectratio]{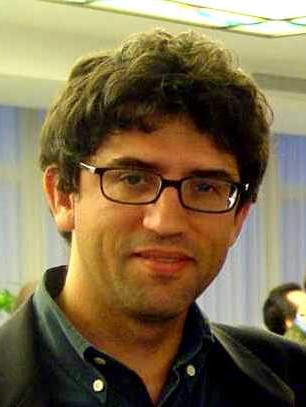}}]
{Giuseppe Caire} (S'92 -- M'94 -- SM'03 -- F'05) 
was born in Torino in 1965. He received the B.Sc. in Electrical Engineering  from Politecnico di Torino in 1990, 
the M.Sc. in Electrical Engineering from Princeton University in 1992, and the Ph.D. from Politecnico di Torino in 1994. 
He has been a post-doctoral research fellow with the European Space Agency (ESTEC, Noordwijk, The Netherlands) in 1994-1995,
Assistant Professor in Telecommunications at the Politecnico di Torino, Associate Professor at the University of Parma, Italy, 
Professor with the Department of Mobile Communications at the Eurecom Institute,  Sophia-Antipolis, France,
a Professor of Electrical Engineering with the Viterbi School of Engineering, University of Southern California, Los Angeles,
and he is currently an Alexander von Humboldt Professor with the Faculty of Electrical Engineering and Computer Science at the
Technical University of Berlin, Germany.

He received the Jack Neubauer Best System Paper Award from the IEEE Vehicular Technology Society in 2003,  the
IEEE Communications Society and Information Theory Society Joint Paper Award in 2004 and in 2011, 
the Okawa Research Award in 2006,   
the Alexander von Humboldt Professorship in 2014, the Vodafone Innovation Prize in 2015, an ERC Advanced Grant in 2018, 
the Leonard G. Abraham Prize for best IEEE JSAC paper in 2019, the IEEE Communications Society Edwin Howard Armstrong Achievement Award in 2020, 
and he is a recipient of the 2021 Leibinz Prize  of the German National Science Foundation (DFG). 
Giuseppe Caire is a Fellow of IEEE since 2005.  He has served in the Board of Governors of the IEEE Information Theory Society from 2004 to 2007,
and as officer from 2008 to 2013. He was President of the IEEE Information Theory Society in 2011. 
His main research interests are in the field of communications theory, information theory, channel and source coding
with particular focus on wireless communications.   

\end{IEEEbiography}

\end{document}

%% file: macros.tex
\long\def\comment#1{}

\newcommand{\Cc}{{\mathcal C}}
\newcommand{\Dc}{{\mathcal D}}

\newcommand{\Sc}{{\mathcal S}}
\newcommand{\Tc}{{\mathcal T}}

\newcommand{\Vc}{{\mathcal V}}

\newcommand{\asf}{{\mathsf a}}
\newcommand{\bsf}{{\mathsf b}}

\newcommand{\Bsf}{{\mathsf B}}

\newcommand{\Ksf}{{\mathsf K}}
\newcommand{\Lsf}{{\mathsf L}}
\newcommand{\Msf}{{\mathsf M}}
\newcommand{\Nsf}{{\mathsf N}}

\newcommand{\Rsf}{{\mathsf R}}

\newtheorem{thm}{Theorem}%[section]

\newtheorem{rem}{Remark}
\newtheorem{example}{Example}

\providecommand{\definitionname}{Definition}